\documentclass[preprint]{aastex}

\usepackage{epsfig,lscape}

\newcommand{\barl}{\mbox{$\bar{l}$}}
\newcommand{\barL}{\mbox{$\bar{L}$}}
\newcommand{\barV}{\mbox{$\bar{V}$}}
\newcommand{\barH}{\mbox{$\bar{H}$}}

\newcommand{\barm}{\mbox{$\bar{m}$}}
\newcommand{\barM}{\mbox{$\bar{M}$}}
\newcommand{\barKs}{\mbox{$\bar{K_s}$}}
\newcommand{\barF}{\mbox{$\overline{F160W}$}}
\newcommand{\barK}{\mbox{$\bar{K}$}}

\newcommand{\barI}{\mbox{$\bar{I}$}}
\newcommand{\barJ}{\mbox{$\bar{J}$}}
\newcommand{\HST}{{\sl HST}}
\newcommand{\samename}{\vrule height0.4pt depth0.0pt width1.0in \thinspace.}

\slugcomment{To appear in ApJ, NN.}
\shorttitle{Surface Brightness Fluctuations}
\shortauthors{Rosa Gonz\'alez et al.}

\received{2003 November 27}
\begin{document}

\newcommand{\hi}{H\,{\sc i}}
\newcommand{\hii}{H\,{\sc ii}}

\title{IR Surface Brightness Fluctuations of 
Magellanic Star Clusters\altaffilmark{1}}

\author{Rosa A.\ Gonz\'alez\altaffilmark{2}}\affil{Centro de Radioastronom\'{\i}a y Astrof\'{\i}sica, UNAM, Campus Morelia, Michoac\'an, M\'exico, C.P.\ 58190}

\author{Michael C.\ Liu\altaffilmark{3}}
\affil{Institute for Astronomy, University of Hawaii, 2680 Woodlawn Drive, Honolulu, HI 96822}

\and

\author{Gustavo Bruzual A.\altaffilmark{4}}
\affil{Centro de Investigaciones de Astronom\'{\i}a, Apartado Postal 264, M\'erida 5101-A, Venezuela} 

\altaffiltext{1}
{This research has made use of the NASA/ IPAC
 Infrared Science Archive, which is operated by the
 Jet Propulsion Laboratory, California Institute of
 Technology, under contract with the National
 Aeronautics and Space Administration.}
\altaffiltext{2}{E-mail address: {\tt r.gonzalez@astrosmo.unam.mx}}
\altaffiltext{3}{E-mail address: {\tt mliu@ifa.hawaii.edu}}
\altaffiltext{4}{E-mail address: {\tt bruzual@cida.ve}}


\begin{abstract}
We present surface brightness fluctuations (SBFs) in the near--IR for 191 
Magellanic star clusters available 
in the Second Incremental and All Sky Data 
releases of the 
Two Micron All Sky Survey (2MASS), 
and compare them with SBFs of Fornax Cluster galaxies 
and with predictions from stellar population models as well. 
We also construct color--magnitude diagrams (CMDs) for these 
clusters using the 2MASS Point Source Catalog (PSC). 
Our goals are twofold. First, to provide an empirical calibration
of near--IR SBFs, given that existing stellar population synthesis
models are particularly discrepant in the near--IR.
Second, whereas most previous SBF studies
have focused on old, metal rich populations, this is the first
application to a system with such a wide range of ages
($\sim$ 10$^6$ to more than 10$^{10}$ yr, i.e., 4 orders
of magnitude), at the same time that the clusters have  
a very narrow range
of metallicities (Z $\sim$ 0.0006 -- 0.01, ie., 1 order of magnitude only).
Since stellar population synthesis
models predict a more complex sensitivity of SBFs to
metallicity and age in the near--IR than in the optical, this
analysis offers a unique way of disentangling the
effects of age and metallicity.


We find a satisfactory agreement between models and data.
We also confirm that near--IR 
fluctuations and fluctuation colors are mostly driven
by age in the Magellanic cluster populations, and that in this respect
they constitute a sequence in which the Fornax Cluster galaxies
fit adequately. Fluctuations are powered by red supergiants
with high--mass precursors in young populations, and by intermediate--mass 
stars populating the asymptotic giant branch in intermediate--age populations. 
For old populations, the trend with age of both fluctuation magnitudes and  
colors can be explained straightforwardly by 
evolution in the structure and morphology of the 
red giant branch.
Moreover, fluctuation colors display a
tendency to redden with age that can be fit by a straight line. 
For the star clusters only,  
(\barH\ - \barKs) = (0.21$\pm$0.03)Log(age/yr) $-$ (1.29$\pm$0.21); 
once galaxies are included, 
(\barH\ - \barKs) = (0.20$\pm$0.02)Log(age/yr) $-$ (1.25$\pm$0.16).
Finally, we use for the 
first time 
a Poissonian approach to establish the error bars of 
fluctuation measurements, instead of the customary Monte Carlo
simulations. 

\end{abstract}


\section{Introduction} \label{intro}

While the mean surface brightness of a galaxy is
independent of distance, the variance about the mean decreases with
distance --- i.e., given the same angular resolution, 
more distant galaxies appear smoother.  
This is the principle behind surface brightness fluctuation 
measurements \citep[SBFs;][]{tonr88,blak01},
one of the most powerful
methods to determine cosmological distances
\citep[e.g.,][]{tonr97,liu01,jens03}.
SBFs arise from
Poisson fluctuations in the number of stars within a resolution
element, and they are 
measured through the observed
ratio of the variance to the mean surface brightness of a galaxy; that is,
the ratio (denoted \barL) of the second to the first moment of the
stellar luminosity function, scaled by the inverse of 
4$\pi d^2$, where $d$ is the distance. SBF
measurements are expressed in \barm\ and \barM, which are,
respectively, the apparent and absolute magnitudes of \barL.

SBF magnitudes, however,
depend not only on galaxy distances, but also
on the age and metallicity of stars. Therefore,
SBFs also offer a unique possibility to investigate unresolved stellar
populations.  For example, as a luminosity--weighted mean, \barM\ is
much more sensitive to giant stars than integrated colors 
\citep{wort93a,ajha94}. For
the same reason, \barM\ is relatively 
insensitive to differences in the IMF for intermediate--age and old
systems.

We engaged in this work with the aim of providing an 
empirical calibration of near--IR SBFs,
specifically for the study of unresolved stellar populations. 
The near--IR is very favorable for SBF measurements,
from the point of view of improved signal (the light of intermediate and
old populations is dominated by the asymptotic giant branch, AGB, 
and the red giant branch, RGB), reduced
dust extinction and, last but not least, the model prediction that  
near--IR SBFs might help break the age--metallicity degeneracy  
\citep{wort93b}. However, there is
the very important disadvantage that 
existing stellar population
synthesis models are particularly discrepant in the near--IR spectral
region \citep{char96,liu00,blak01}.
The disagreement is as high as $\sim$ 0.2 mag in $(V\!-\!K)$, compared to
$\sim$ 0.05 mag in $(B\!-\!V)$. 
The ill--determined contribution of asymptotic giant branch (AGB) 
stars to the integrated light may
be the most important source of this problem \citep{ferr95}. 
Such an uncertainty is bound to compromise the calibration of \barM.
{\it An empirical calibration of near--IR SBFs is therefore essential}.

\section{Our strategy} \label{strategy}

In theory, a good starting point to assess the impact
of stellar population variations
on SBFs, and to empirically calibrate \barM\ would be to derive
fluctuations for a number of
simple stellar populations with known distances.  
This is not a new idea.
\citet{ajha94} attempted to calibrate the SBFs
zero--point with $V$ and $I$ photometry of Galactic globular clusters.
Their derived \barM$_I$, however, does not constrain \barM$_I$ for galaxies,
since the range of ages of Galactic globular clusters is small,
and in general their metallicities do not overlap with 
those of spiral bulges
and early--type galaxies.  About five \citep{harr96}  
of the inner--disk subgroup of the
Galactic globular clusters have metallicities 
in the right range; aside from the fact that ages and distances
of those clusters are generally not very well determined yet, 
their number is so small that their
analysis will be dominated by stochastic effects (see \S \ref{errors}). 
Finally, \citet{ajha94} found that optical data alone are inadequate
to decouple the effects of age and metallicity reliably.

A study of the Magellanic Clouds (MC) clusters 
would seem to constitute a better course of action from the
point of view of their relevance to early--type galaxies, 
for two reasons:
the clusters have very well known distances, and 
they span a much wider range of ages 
($\sim 10^6$ to $\sim 10^{10}$ yr) 
than the Galactic globular clusters
(all with $\sim 10^{10}$ yr).
The oldest MC clusters are as old or older than elliptical galaxies
and spiral bulges.
On the other hand, it is true that their metallicity is low
(Z $\sim$ 0.0006 -- 0.01), but  
their slow chemical enrichment history means that
clusters with ages between a few Myr and 3 Gyr have all basically 
Z $\sim$ 0.01 \citep{cohe82}. 
Since stellar population synthesis models predict a
more complex sensitivity to 
metallicity and age in the near--IR than in the optical 
SBFs \citep{wort93a,liu00}, a near--IR study of the MC 
star clusters, with their wide range of ages (4 orders of
magnitude) and narrow range
of metallicities (1 order of magnitude only), 
could offer a unique way of disentangling the effects of
age and metallicity. 

In reality, for this approach to work, the sample should include
as many clusters as possible, because it is in star clusters 
where the AGB problem manifests itself most dramatically.
In each individual cluster, the stars populating the
AGB and the upper red giant branch (RGB) are so few that they
do not properly
represent the distribution of the brightest
AGB/RGB stars
on the isochrone. Often, the integrated near--IR light and,
even worse, the SBFs will be dominated by a single
luminous, cool star.
The way around this problem is an appropriate
treatment of a sufficiently rich database. Fortunately,
$J$, $H$, and $K_s$ data of all MC clusters are now  
available in  
the Two Micron All Sky Survey \citep[2MASS;][]{skru97}.
Rather than analyzing each cluster separately, in order to 
reduce stochastic effects we have
built ``superclusters,'' by coadding clusters
in the \citet{elso85,elso88}
sample that have the same 
SWB class \citep{sear80}.
The SWB classification is based on two 
reddening--free parameters, derived from integrated 
$ugvr$ photometry 
of 61 rich star clusters in the Magellanic Clouds; it 
constitutes a smooth, one--dimensional sequence of increasing age and
decreasing metallicity. \citet{elso85} assigned SWB classes to
147 more clusters using $UBV$ photometry, assuming a low 
and uniform reddening ($E_{B - V} \approx$ 0.1 $\pm$ 0.1) towards and 
in the clouds, which is valid in general for 
clusters older than a few times 10$^7$ years \citep{char00}. 
The \citet{elso85} classification is parameterized by $s$,
where $s$ = (5.75 $\pm$ 0.26) SWB class + (9.54 $\pm$ 1.45). 
We have grouped the clusters in superclusters according to their
$s$-parameter, rather than their SWB class, as shown in
Table \ref{tabclust}.
Nevertheless, we have assigned ages and metallicities to the
superclusters from \citet{cohe82}, by virtue of their 
SWB types.\footnote{The exception is the Pre-SWB supercluster,
for which we have assumed the age of a cluster with 
$s$=7, or the ``central'' $s$-type of its constituents.} 

In order to compare the results obtained for star clusters with
those for galaxies, we use \barM$_{K_s}$ 
and \barM$_{F160W}$ derived for
a sample of Fornax Cluster galaxies by, respectively,
\citet{liu02} and \citet{jens03}. 
Likewise, we have
taken the $J$, $H$, and $K_s$ integrated fluxes and colors of the Fornax
galaxies directly from the 2MASS Second Incremental Release Extended
Source Catalog (XSC), via the GATOR catalog web query page.
Finally, ages and abundances have been adopted from
\citet{kunt98}. Parameter values for the Fornax Cluster galaxies
are all presented in Table \ref{tabfornax}.

We had to convert \barM$_{F160W}$ of the galaxies to \barM$_H$.
From the photometric transformations between the  
\HST\ NICMOS Camera 2 filters and the CIT/CTIO system,
published by \citet{step00}
for cool giants of near-solar metallicity, we get:


\begin{equation}
m_H = m_{F160W} + (0.080 \pm 0.069) - (0.243 \pm 0.046) (m_J - m_K). 
\label{etransf}
\end{equation}

\noindent
However, the transformation coefficients for fluctuation magnitudes
may be different from these,
since the spectrum of the fluctuations is not the same as the spectra of
the stars used to derive the transformation. \citet{buzz93} 
--for transformations of SBF magnitudes between Johnson $R, I$, 
and Cousins $R, I$-- and \citet{blak01} --for transformations between
$V_{F555W}$, $I_{F814W}$, and Johnson $V$, Cousins $I$--
maintain that transformation equations obtained from stellar 
observations can be used for fluctuation magnitudes of galaxies if 
fluctuation colors are substituted for integrated ones. 
On the other hand, \citet{tonr97} state that, instead, the mean 
($V - I$) color of the fluctuations --which these authors use to 
correct their SBF magnitudes-- is the mean of the integrated 
color and the fluctuation color (\barV\ - \barI).
In view of these conflicting statements, we have used models
(see \S \ref{errors}) to decide which color is most appropriate
to use in the above equation to transform \barF to \barH. For 
model populations with solar metallicity and Z = 0.05 --the 
most relevant metallicities for the galaxies--, we have plotted 
in Figure \ref{ftransf} (\barH\ - \barF) vs ($J - K$), (\barJ\ - \barK),
and integrated ($J - K$)$_{bright}$ of stars brighter than $M_K \leq$ -4.5
(which are the ones we can detect as resolved sources at the
distance of the LMC and which mostly determine the 
fluctuation values, see \S \ref{data}). Model transformations are 
closest to equation \ref{etransf} when using the 
color of the brightest stars, which also
turns out to be approximately the mean of the integrated ($J - K$)
and (\barJ\ - \barK), as described by \citet{tonr97}. 
Since we do not have $K$ for the galaxies, but $K_s$,
we have also used the models to check that 
the uncertainty introduced by using $K_s$
is smaller than the systematic error made when applying the stellar
transformation to the \barF\ measurements.
Of course, we cannot measure ($J - K_s$)$_{bright}$ directly but,
from the models, 
it is approximately 0.48 mag redder than the integrated 
($J - K_s$) for the color range of the galaxies (0.85 -- 0.95).  
(Unfortunately, we do not have (\barJ\ - \barKs) for 
the galaxy data, so we cannot derive empirically the mean ($J - K_s$)
color of the fluctuations, either.) 
Summarizing, we use the following transformation equation:

\begin{equation}
\barM_H = \barM_{F160W} + (0.08 \pm 0.07) - (0.24 \pm 0.05) (J - K_s + 0.48). 
\end{equation}

\section{Treatment of the data} \label{data}


As stated above, \barL\
is the ratio of the second moment of the luminosity function
to its first moment, the integrated luminosity.
This can be expressed with the 
following equation: 

\begin{equation}
\barL \equiv \frac{\Sigma n_i{L_i}^2}{\Sigma n_i L_i}, 
\label{eqsbf}
\end{equation}

\noindent where $n_i$ is the number of stars of 
type $i$ and luminosity $L_i$. 
Bright stars are the main contributors
to the numerator, while faint stars contribute significantly
to the denominator.
In contrast to the measurements performed in distant galaxies,
where pixel-to-pixel SBFs probe unresolved stellar populations, in star 
clusters the second moment of the stellar luminosity function, 
or the numerator, is derived from
measurements of resolved, bright stars \citep{ajha94}. 
The integrated luminosity or the denominator, on the other hand, 
is equal to the total light detected in the
image, after removal of any sky background emission.
The depth of the 2MASS survey is more than adequate for our 
purposes. In the optical and at least for 
Galactic halo globulars, the second moment of the 
luminosity converges quickly,
with 99\% of the sum being obtained with the three brightest
magnitudes of cluster stars \citep{ajha94}.
The southern 2MASS survey was carried out with a 1.3-m telescope 
at the Cerro Tololo Inter-American Observatory (CTIO); the 
$J$, $H$, and $K_s$ data were
secured simultaneously with a 3-channel camera, equipped with 
three 256$\times$256 NICMOS3 arrays.   
The seeing throughout the 2MASS observations ranged ---respectively for
$J$, $H$, and $K_s$--- from 2\farcs4 to 5\farcs4, 2\farcs5 to
5\farcs0, and 2\farcs5 to 4\farcs6;
the average seeing was 2\farcs8 for
$J$ and $K_s$, and 2\farcs7 for $H$.\footnote{http://spider.ipac.caltech.edu/staff/roc/2mass/seeing/seesum.html}
The raw camera pixel size of the survey was 2\arcsec, 
but the sampling was improved by dithering. The released 
data have 1\arcsec\ pixels, and we have measured
an average FWHM of 3\arcsec, in all three bands. 
Nominally, the 2MASS PSC is 100\%  
complete for $K_s <$ 15 mag at the general position of 
the Magellanic Clouds.\footnote{http://www.ipac.caltech.edu/2mass/releases/allsky/doc/sec2\_2.html} 
To err on the conservative side, though, 
we have inspected the luminosity functions 
of the clusters within 1\arcmin\ of their centers. Eighty seven per cent 
of them seem complete for $K_s <$ 14 mag; 
this means that   
the four brightest magnitudes of stars at the
distance of the Large Magellanic Cloud (LMC) have 
been detected (see Fig.\ \ref{cms}). 
In \S \ref{errors} below, we demonstrate
that these should suffice for the
calculation of fluctuation magnitudes and colors in the
near--IR.

We retrieved from the 2MASS archive $J$, $H$, and $K_s$  
data for 156 MC clusters with an $s$-parameter
\citep{elso85,elso88}, and that were available in the  
Second Incremental Release of the 2MASS 
database as Atlas (i.e.,  
uncompressed) images. Later, we obtained images
restored from lossy--compressed files  
for another 35 clusters from the All Sky release.
The data were then used to build eight ``superclusters,''
one for each of the seven different SWB classes \citep{sear80}, 
plus one ``Pre-SWB-class'' supercluster.
Besides the images, photometry for the point sources 
was obtained from the 2MASS Second Incremental Release 
and All--Sky PSCs; 
the coordinates used to retrieve the 
source lists were  
mostly those provided by SIMBAD, although
for many of the Small Magellanic Cloud (SMC) objects 
we used the coordinates in 
\citet{welc91} and, occasionally, positions determined 
from the $J$ images by eye. 
At this stage, the $J$ photometry was 
used to find a center of light for all the images.  

Once centered, we proceeded to assemble the supercluster mosaics  
by coadding the individual clusters of each SWB class, after 
subtracting the sky value registered in the image header,\footnote{
For several of the clusters, we confirmed the accuracy of this value
by eye, by looking at the mode in an annulus between 2\farcm0 and 
2\farcm5 from the cluster center, and with DAOPHOT
\citep{stet87}.} 
multiplicatively scaling each one to a common photometric 
zero-point, dereddening (even though reddening is not very
important at these wavelengths), and geometrically magnifying SMC clusters  
to place all of them at the same distance modulus of the 
LMC. We take 
$(m - M)_{\rm o} = 18.50 \pm 0.13$ for the LMC, and 
$(m - M)_{\rm o} = 18.99 \pm 0.05$ for the SMC,\footnote{The errors listed in
Tables \ref{tabresI} and \ref{tabresII}  
for the derived absolute integrated and fluctuation magnitudes of the 
MC superclusters do 
not include the quoted errors in the clouds' distance
moduli, nor dispersions due to their depths along the 
line of sight. Neither should affect the conclusions of this study.
The {\it systematic} error in the \HST\ Cepheid
distance to the LMC, estimated to be about $\pm$ 0.16 mag 
\citep{moul00}, is also not taken into account;
future adjustments to the LMC distance modulus would result 
in a constant offset applied to the SBF absolute magnitudes of the 
Fornax galaxies we have taken from the literature and of the MC superclusters 
that we derive here.} after the Cepheid distances in 
\citet{ferr00}. We also masked out bad columns. 
An important final step was to correct for any residual over 
or undersubtraction of the sky emission, by measuring the background 
of each supercluster mosaic in an annulus between 2\farcm0 and 2\farcm5
from the center. 
The mosaics were used mainly to measure the integrated light of
the superclusters, a quantity that goes  
into the denominator of the expression for \barL. With the aid of
a few star clusters for which both uncompressed and compressed 
images were available, we checked that the compressed ones were
adequate to perform these measurements, for the few cases where 
only the latter were released. 
We also inspected the radial profiles of the mosaics 
in order to look for anomalies 
(e.g. extremely bright, probably foreground, stars) and 
to assess the contribution to the integrated light from different
radii. 

Next, we went back to the PSC in order to assemble star lists
for each supercluster.\footnote{
For several of the individual star clusters, we verified the
accuracy of the published photometry with DAOPHOT. As an
external check, the near--IR photometry obtained by \citet{ferr95} for 
12 globular clusters in the Magellanic Clouds is virtually identical 
to their PSC raw (i.e., dismissing quality flags) photometry; 
when using the flags to eliminate sources, the CMDs produced with 
the PSC values are tighter than those published by Ferraro and 
collaborators. These authors obtained their data with the
1.5-m telescope at CTIO and an InSb array, 58 $\times$ 62 pixel, 
with a pixel size of 0\farcs92. It is perhaps worth mentioning that the 
similarity between the PSC and the \citet{ferr95} results 
is what convinced us
in the first place that this project could be done with the 2MASS
data. 
} 
This time, the star lists for the 
individual clusters were retrieved with VizieR \citep{ochs00} using the 
centers determined from the $J$ light centroids. Afterwards, the 
distance of each star from the center of the supercluster 
was adjusted to account for the differences in distance 
among the clouds, again to place all the stars at 
a distance modulus of $(m - M)_{\rm o} =$ 18.50; the photometry, too,
was corrected for the differences in distance, as well as 
for reddening. Finally, sources with dubious photometry were
eliminated, as were outliers. To evaluate the quality of
the photometry, we used the flags from the PSC itself. 
We kept only sources within 1\arcmin\ from the 
centers of the superclusters; that had been detected and had no
artifacts in all three bands; that  
had read--out--2 -- read--out--1 profile--fit
photometry,\footnote{
The integration 
time for each frame in the 2MASS survey included: two 51 ms resets, 
one 51 ms ``Read\_1" (R1) integration, and one
1.3 s ``Read\_2" (R2) integration. An additional 
delay of 5
ms was added to allow for overhead and settling.
2MASS Atlas images were produced by the
coaddition of six overlapping R1 -- R2 frames, each with 1.3 s 
integration, for a total integration time of 
7.8 s. Point sources were 
detected from the Atlas images, but the position
and photometry of faint sources 
(most sources in the catalog) were estimated 
through profile--fitting in each of the 
six stacked R1 -- R2 frames \citep{cutr03}.}
also at $J$, $H$, and $K_s$; and that were 
not associated with either an extended source, a minor planet, or a 
comet. In order to minimize field contamination, we followed the criterion 
of \citet{ferr95}, that is, we excluded from the analysis stars in the range 
12.3 $< (K_s)_o <$ 14.3 with colors ($J - K_s$)$_o >$ 1.2 or 
($J - K_s$)$_o <$ 0.4. Given that the radius of 1 arcmin\ 
encompasses only the centers of the star clusters,    
no further field decontamination scheme was applied.

Out of these starlists, three CMDs were produced 
for each supercluster, one for stars with $r < 0\farcm34$, 
a second one for stars with $0\farcm34 \leq r < 0\farcm66$ and,
finally, another one for stars with $0\farcm67 \leq r \leq 1\farcm0$. 
The comparison between the 3 diagrams of each supercluster 
reassured us that field contamination was not a problem.
We got rid of probable 
foreground stars: one extremely bright star in
NGC~1754 (SWB VI), another in NGC~1786 (SWB VII), and a
couple of bright, blue, stars in NGC~1777 (SWB V).  

Lastly, integrated fluxes, integrated colors,  
absolute fluctuation magnitudes, and fluctuation colors were obtained
for each supercluster. 
The integrated fluxes, which are also needed for the
denominator of eq.\ \ref{eqsbf}, were acquired simply by 
summing up the flux in all the pixels within 1\arcmin\ of the center 
of each supercluster, and subtracting the flux from foreground 
stars and from bright stars (i.e., sources in the PSC catalog with 
$K_s \leq$14 at the distance of the LMC) 
that had dubious photometry, as judged from the PSC flags; 
this is mathematically analogous, 
but procedurally much easier, than measuring each separate cluster image  
and then adding.
The numerator of eq.\ \ref{eqsbf}, on the other hand, was calculated  
by performing sums over the individual stars with good photometry in the same 
region. 

We present in Figure \ref{mosaics} images of all eight superclusters.
These are greyscale versions of $J$, $H$, and $K_s$ color mosaics.
Figure \ref{cms} displays,  
again for all superclusters,
the CMDs of the stars within 1\arcmin\ of their centers.
The average photometric errors are 0.04 mag 
in brightness and 
0.02 mag in color for sources with 
$K_s \leq$ 13; respectively, 0.06 and 0.03 mag 
for stars with 13 $< K_s \leq$ 14; and 0.13 and 0.07 mag 
(about the size of the dots) for sources with 14 $< K_s \leq$ 15. 
The width of several of the diagrams (conspicuously, II and III) results 
from the fact that our artificial clusters are not fully
homogeneous populations; unfortunately, binning the 
data in superclusters is the compromise we  
have found most convenient to adopt in order to try to circumvent 
the problem of small number statistics
posed by individual star clusters (c.f.\ \S \ref{strategy}
and \S \ref{errors}). As we will see below (\S \ref{results}),
this approach lets the general tendencies of the data
show through, while hopefully eliminating possible biases
and reducing random errors.  

Table \ref{tabclust} lists assumed age and metallicity 
of the superclusters; limiting radii of
analyzed regions (in arcseconds {\em at the distance of the LMC}); star 
clusters that went
into building each supercluster; 
number of stars of each individual cluster that contributed to the
calculation of the supercluster SBFs; 
$s$-parameter from 
\citet{elso85} or \citet{elso88}; whether the star clusters belong to the 
LMC or the SMC; 
$E(B-V)$ from \citet{pers83}; and SWB-class from
\citet{sear80}. 
When clusters do not have individually measured reddening, we
have assumed $E(B-V) = 0.075$ for the LMC and $E(B-V)=0.037$ for the
SMC \citep{schl98}; also from \citet{schl98}, we have taken
$A_J = 0.902 E(B-V)$, $A_H = 0.576 E(B-V)$, and 
$A_K = 0.367 E(B-V)$.

%

\section{Models and errors} \label{errors}


We use the most recent \citet{bruz03} 
evolutionary stellar population synthesis models.
Here, we utilize what these authors call the
``standard'' reference models for different 
metallicities, built using the 
Padova 1994 stellar evolution isochrones
\citep{alon93,bres93,fa94a,fa94b,gira96}; 
the model atmospheres compiled by \citet{leje97,leje98}, 
as corrected by \citet{west01} and 
\citet{west02}; and the IMF parameterized by 
\citet{chab03}, truncated at 0.1 $M_{\odot}$ and 100 $M_{\odot}$.
Our choice of the ``standard'' models is based on 
the assessment of \citet{bruz03}, and on tests that we made 
specifically with the MC cluster data as well.

We compute SBF and integrated magnitudes
and colors, through the 2MASS near--IR filters, 
of single--burst stellar populations for ages
of 1 Myr -- 17 Gyr and metallicities $Z$ = 0.0004 -- 0.05.
 
First of all, we have used the models to check that, in fact,
the stars that are detected as point sources by the 
2MASS survey are enough to obtain a reliable estimate of
the near--IR SBFs of the Magellanic star clusters.  
Fig.\ \ref{convtest} shows, for Z = 0.0004,
0.004, 0.008, and 0.05, 
the difference between the $J$, $H$, and $K_s$ 
integrated and fluctuation magnitudes
calculated with all the stars and only from those with $M_{K_s} \leq$ -4.5
(or $K_s$ = 14 at the LMC).\footnote{In order to obtain the
fluctuation magnitudes, the
second moments derived from the bright stars are
still normalized by the integrated luminosity of all the
stars, as is done with the data.}
This difference is actually an overestimate, since we do detect fainter stars.
Excepting extremely young ages that are not relevant to
this work,
even when the contribution from the bright stars to the
integrated luminosity is of the order of 10--20 percent, at
an age of half to 1 Gyr, depending on 
metallicity ---when supergiants have died and
giants are yet to appear---, or at $\sim$ 10 
Gyr ---when 
the luminosity of the RGB dwindles (see \S \ref{disc})---, 
the foreseen differences in
the derived fluctuation magnitudes are
always smaller than the expected empirical
errors due to stochastic fluctuations in the number of stars
(cf.\ Fig. \ref{convtest} and Table
\ref{tabresII}), as we discuss here below. 
As an additional check, Table \ref{tabfrac} compares, 
for all eight superclusters, the 
theoretical and measured contributions from stars with $M_{K_s} \leq$ -4.5 
to the {\it integrated} $J$, $H$, and $K_s$ fluxes. 
There is quite a good agreement, except for the intermediate
age superclusters classes III and IV, where the models underestimate
the light fraction from bright stars by a factor of $\sim$ 2.\footnote{
Although models with higher metallicity also predict a larger 
contribution from bright stars, even models with Z = 0.05 
cannot match the data.}
The actual minimum contributions of 
$\sim$ 20\% are observed for class VII 
(remarkably, exactly at the levels 
anticipated by the models); 
these minimal ratios
would translate into discrepancies in the 
SBFs that would increase the 
errors quoted in Table \ref{tabresII} by a few 
hundredths of a magnitude only.

\subsection{Stochastic errors} \label{mycontrib}

Stochastic errors due to small number statistics are central to 
SBF studies of star clusters.
The standard way to assign error bars to fluctuation magnitudes and 
colors has so far been through Monte Carlo simulations
\citep[e.g.,][]{ajha94,bruz02}. Recently, however,
\citet{cerv02} have presented an approximate statistical formalism to
estimate quantitatively the dispersion expected in 
relevant observables of simple stellar populations, owing 
to statistical fluctuations of the luminosity function. 
This approach is based on the assumption that 
the variables involved have a Poissonian nature; it 
shows explicitly that models can be used to accurately 
predict the scatter observed in real data, if the 
theoretical relative error produced by stochastic
fluctuations in the number of stars is scaled by 
$M_{tot}^{-1/2}$, where $M_{tot}$ is the total 
mass of the stellar population.
An important ingredient when calculating the dispersion 
in this fashion are the covariance terms.
The covariance between two quantities,
cov$(x,y)$, is defined as 

\begin{equation}
{\rm cov}(x,y) = \rho(x,y) \sigma_x \sigma_y,
\end{equation}

\noindent
where $\rho(x,y)$ is the correlation coefficient and
$\sigma_x$ and $\sigma_y$ are, respectively, the uncertainties
due to small number statistics in $x$ and $y$.
$\rho(x,y)$ varies 
between -1 and 1, depending on the sign of the 
correlation: a positive correlation means that the quantities
vary together in the same direction; a negative correlation
occurs when the quantities vary together in opposite directions
(in this work, we will denote with a minus sign a covariance with
a negative correlation coefficient).
If $\rho(x,y) =$ 0, the quantities are not correlated, i.e., 
they are statistically independent. For one given star,
for example, the luminosities in different bands are completely
correlated. For a group of stars, though, the contribution from
each star has to be considered for a proper calculation of
the covariance. In general, in the case of 
observables ---like integrated colors and surface brightness 
fluctuations of stellar populations--- that are ratios of luminosities, the 
adoption {\it a priori} of $\rho =$ 0 will 
overestimate the error.

\citet{cerv02} verify that their method works  
for integrated properties, such as colors and equivalent widths of
emission and absorption lines, through  
cross--checks with the outcome of, again, Monte Carlo simulations.
Here, we apply the method of \citet{cerv02} to derive
error bars for integrated magnitudes, integrated 
colors, fluctuation magnitudes, and fluctuation colors.
The operations performed in each case are presented
in Appendix \ref{errprop}. 
Through new comparisons with Monte Carlo simulations, 
we have corroborated that the chosen approach 
is applicable to fluctuation magnitudes and colors as well.
For cluster masses higher than a few$\times 10^5$ $M_\odot$, 
our ``analytical'' error bars and the Monte Carlo 
errors are equivalent, although one must bear in mind that
these analytical errors are 1-$\sigma$, whereas  
Monte Carlo simulations will generate results 
within 3-$\sigma$ of the ``central'' analytical value.\footnote{  
For lower cluster masses, sampling fluctuations will 
produce biases (i.e., deviations from the 
theoretically predicted values of the 
observables) and multimodality that depend on 
wavelength and stellar evolutionary phase, and that cannot be 
accounted for by our analytical calculation of the 
stochastic errors \citep{sant97,cant03,cerv03,
raim03}. Investigating these effects is beyond the
scope of this paper.}

Furthermore,
we have calculated errors not just from the models, 
but also directly from the data. 
When working with the models,
all equations are applied as written down in the appendix; 
there, $w_i$ stands for the number of stars of mass $m_i$ by unit mass,
and $M$ is in each case the mass of the 
supercluster in question. The supercluster masses are
obtained from the models themselves, 
using the theoretical near--IR mass--to--light
ratios of a population 
with the same age and metallicity as each supercluster; the 
tabulated errors in the masses in Tables \ref{tabresI} and 
\ref{tabresII} are equal to the dispersion
of the results at $J$, $H$, and $K_s$.\footnote{
The supercluster mass increases with 
age; this is probably just a manifestation that 
more massive clusters are more resilient against disruption  
and survive longer.} 
When dealing with data, however, we perform sums 
over individual stars; hence, all the
$w_i$'s are assumed to be unity. Also, in the case of
the data, it is unnecessary to normalize by $M^{-1/2}$,
since this operation is implicitly done by adding the stars. 
Typically, the errors derived from the data 
---which are the ones we quote--- 
are about 2--3 times the size of the error bars
anticipated from the models.\footnote{Exceptionally,
the data errors for the youngest (Pre-SWB) supercluster 
are $\sim$ one order of magnitude larger than those
predicted by the models. In this case, the mass of 
the supercluster
could have been overestimated; it probably is no coincidence that
this is the mass determination with the largest (absolute and relative) 
error (see Tables \ref{tabresI} and \ref{tabresII})} 

Yet another check has been provided by the comparison 
between our errors and those quoted  
by \citet{liu02} for \barKs\ and 
($V - I_c$) of Fornax Cluster galaxies.
The errors in the present paper are 
about twice larger, for both integrated and
fluctuation magnitudes. 
Since the Fornax Cluster elliptical and S0 galaxies analyzed
are 3--4 orders of magnitude more massive than
our superclusters,\footnote{
From their line--of--sight velocity dispersions
(Hypercat, http://www-obs.univ-lyon1.fr/hypercat,
which uses an updated version of the literature compilation by
\citet{prug96}; \citet{grah98})  
and assuming a mass--to--light ratio of $\sim$ 2 \citep{fabe76},
the galaxies have masses $\leq$ 10$^9$ M$_{\odot}$.} 
and given that relative errors scale as
$M^{-1/2}$, it follows   
that the stochastic effects are 
about 30 times smaller for the Fornax galaxies, and hence a negligible
contributor to the error budget of their derived 
SBFs. This budget includes contributions from distance errors,
the depth of the Fornax cluster and, necessarily, the 
composite nature of the stellar populations of the galaxies. 

Incidentally, we have calculated 
approximately the (unknown) correlation coefficients 
$\rho(a,b)$ for model integrated and 
fluctuation near--IR colors, assuming that the
colors have the form $u = a/b$, where 

\begin{equation}
\frac{\sigma^2_u}{u^2} = \frac{\sigma^2_a}{a^2} + \frac{\sigma^2_b}{b^2} 
- 2 \frac{{\rm cov}(a,b)}{ab}.  
\end{equation}

A comparison 
between the results obtained with the appropriate ``exact'' calculation 
in Appendix \ref{errprop}, and the ``na\"{\i}ve" formula above
yields $\rho(a,b)$. We find that, for all metallicities, 
the coefficients $\rho(a,b)$ for fluctuation colors 
are $\sim$ -1 at all ages,
while those for integrated colors are $\sim$ -0.9 in the case of a  
very young population, but reach $\sim$ -0.98 after about 3.5 Myr.  


\subsection{Systematic errors} \label{syserr}

Aside from the problem posed by small number statistics, 
two other issues are crucial for SBF 
measurements of star clusters, in view of the way they  
are performed: the sky level, which impacts the denominator
of equation \ref{eqsbf},
and crowding which, through the blending 
of sources, will in principle make the numerator larger and
hence the SBF magnitude brighter.
We invite the reader here to inspect the top left 
panel of Figure \ref{figsyst}, where we present
the $H$-band absolute fluctuation magnitude vs age 
of the eight MC superclusters, carried out as we 
have described in \S \ref{data}. Models of different metallicities, 
that will be discussed in detail in \S \ref{flucmag}, are also plotted.
The color--coded dots are measurements obtained from different
regions of the superclusters:  the central 0\arcsec\ to 20\arcsec\ 
(red), the annulus between 20\arcsec\ to 40\arcsec\ (blue), and 
the annulus between 40\arcsec\ to 60\arcsec\ 
(green); the black dots are the values derived from the analysis 
of the whole region within 1\arcmin\ of the center of the 
superclusters. The central red dots should be most affected by
crowding, while the outside green dots should be most sensitive
to over or undersubtraction of sky emission.

First, we notice the paradoxical result that, with the exception of
the youngest Pre-SWB supercluster, the red dots are fainter than 
the rest, instead of brighter. While the centralmost region 
of the Pre-SWB supercluster 
shows the expected bias, the remaining superclusters display an indirect
effect of crowding, by virtue of which blended sources appear (rather than
brighter) as suffering from bad photometric measurements, and hence 
are discarded. The green dots derived from the outermost annuli, on the
other hand (and again with the exception of the Pre-SWB supercluster),
tend to appear a little too bright, presumably as a consequence of sky emission
oversubtraction. However, excepting the 
central regions of classes V and VI, the results for all the regions 
of all the clusters are consistent with each other. 
Moreover, on the question of crowding, \citet{ajha94} discarded 
for this reason regions of their data where the two brightest magnitudes covered 
more than 2\% of the area. We have inspected the luminosity functions of 
the individual MC star clusters (including all sources in 
the PSC, regardless of photometric quality) and found that, while the regions 
within 20\arcsec\ of the centers of most of them would be deemed 
crowded by this criterion, already the regions within 40\arcsec\ 
of the centers, taken as a whole, would not.   
Consequently, we will
adopt the measurements from the circular regions within 
1\arcmin\ of the center of the superclusters, given that they seem 
to provide the better balance of uncertainties owing to crowding and 
sky subtraction, and that (unlike the annuli between 
20\arcsec\ and 40\arcsec) they should also be the least  
affected by small number statistics. 
 
The middle left panel of Figure \ref{figsyst} shows the 
($\barH\ - \barKs$) fluctuation color vs age for the MC superclusters. 
Measurements obtained from different regions are color--coded as
before. Although the relative position of the different values 
obtained for each cluster are harder to interpret in terms of 
crowding or faulty subtraction of sky emission, we find again that 
all the results are consistent within the Poisson error bars. 

We show in the bottom left panel of Figure \ref{figsyst}
the ratio of the observed and model contributions of 
stars brighter than $M_{K_s}$ = -4.5 to integrated light
at $H$. 
Here we see once more that blended stars have been preferentially
thrown out in central regions, while outermost annuli 
are very sensitive to sky emission subtraction.
For superclusters I, II, V, VI, and VI, not only 
the match between models and data within  
1\arcmin\ of their centers is good, but the color points scatter
around 1. However, for superclusters Pre, III, and IV, all points 
lie above unity. Especially in the case of classes III and IV,
this is likely pointing to a lack of theorical  
understanding of AGB stars and their
contribution to the integrated light of the clusters, 
rather than caused by problems with the data and their treatment.

For completeness, in the right panels of Figure \ref{figsyst} 
we present the same measurements, but now including the fluxes from 
all point sources, irrespective of their photometric quality. 
The agreement with the 
models is of the same order as before and, if drawn from these 
results, our conclusions would hold.

\subsection{Other errors} \label{othererr}  

We have assumed that the photometric error in the 
integrated luminosity of each individual cluster 
frame is at most 10\%, or the 2MASS survey specification 
for galaxy photometry.\footnote{http://pegasus.phast.umass.edu/2MASS/teaminfo/level1.ps} Given that between 12 and 35 individual 
clusters have been combined to construct each supercluster 
(Tables \ref{tabresI} and \ref{tabresII}), added in 
quadrature these errors would translate
into a photometric uncertainty of 0.02 -- 0.03 mag in  
the integrated luminosity of the superclusters, at which 
point we should be hitting systematic effects.
The photometric errors of the fluctuation 
luminosities and colors have contributions from both the 
integrated and the point source photometry. 
However, in view of the errors in
the point source photometry quoted by the 2MASS PSC,
the total photometric errors are dominated by the 
aforementioned uncertainty in the integrated luminosity. 
Accordingly, and hopefully conservatively, 
we adopt a photometric error of 0.03 mag in the superclusters'
integrated luminosities, integrated colors, fluctuation luminosities 
and fluctuation colors, which we add in quadrature to the 
stochastic errors discussed above in section \S \ref{errors}.
  
In the case of Fornax Cluster galaxies, errors for 
\barKs\ fluctuation magnitudes are
taken from \citet{liu02}, and those for 
\barH\ come from \citet{jens03}. 
The uncertainties in their fluctuation colors are computed assuming
a correlation coefficient $\rho$ of -0.9. 
Errors in integrated colors of galaxies are
calculated from the uncertainties in their integrated fluxes 
quoted by the 2MASS Extended Source Catalog, also with 
$\rho=$ -0.9. The uncertainties 
in the galaxy ages and abundances, [Fe/H] (transformed to errors in 
metallicity, Z), are averages taken from Figure 12 in \citet{kunt00}; 
they both have the value of $\pm$0.15 dex. Finally, 
for the clusters and after \citet{cohe82},
we take errors of $\pm$0.2 dex in abundance, 
and uncertainties in their ages 
of $\pm$0.3 dex. 

\section{Results} \label{results}

Our measurements of MC superclusters are summarized 
in Tables \ref{tabresI} and 
\ref{tabresII}. There we give, respectively, 
cluster integrated and fluctuation magnitudes and 
colors. The results proper are presented in Figures \ref{figcol} 
through \ref{flcolvsz}. 

\subsection{Integrated colors vs age} \label{intcol}

In order to put the SBF measurements in context, we begin 
by plotting in Figure \ref{figcol} the integrated colors 
vs log (age). The left panels overplot the Fornax and MC
data points with the models, while the right panels 
display the data points with error bars. 
Models redden with age, first as massive stars 
become red supergiants, then as the AGB is populated by 
intermediate mass stars, and finally as the brighter 
stars in the populations are low mass stars 
in the RGB and AGB. 
After 10$^7$ years,
models of the same age with higher metallicities are redder, 
owing to increased opacity, which causes the stellar
atmospheres to expand and cool. For stars with 
$T_{eff} \approx$ 3000 -- 6000 K, the main source of opacity 
are the H$^-$ ions, for which metals with low--ionization
potential are the principal
electron donors; in addition, opacity from 
lines and molecular
bands (depending on stellar temperature) also increases 
with higher metallicity.

Focusing on the data, we notice that the Pre-SWB
supercluster is noticeably redder than the models in all three colors, 
$\Delta$($J - H$) and $\Delta$($H - K_s$) $\sim$ 0.4. 
This is probably due mainly to the fact that the age
we have assumed for the supercluster is only an average 
---the supercluster includes objects from 10$^6$ to almost 10$^7$
years old---, and at this point in their evolution clusters 
redden rapidly as they age. Dust reddening might also 
have a (very small) role, since younger clusters 
suffer on average from 3 times more extinction than 
older ones \citep{char00}. The extreme discrepancy between bluer models
and redder clusters decreases gradually through class I and 
all but disappears by class II, at a few $\times$ 
10$^7$ years. According to the models, for objects older 
than a few $\times$ 10$^8$ years 
it is hard to use integrated colors to discriminate
between metal--rich and metal--poor objects. However, the two oldest
superclusters, classes VI and VII, fall exactly
in the loci predicted by the models given their 
metallicities. 
  

\subsection{Fluctuation magnitudes vs age} \label{flucmag} 

We get a remarkable general agreement between 
models and data,
including the Fornax Cluster galaxies,
when we look at fluctuation magnitude vs log (age), Fig.\ \ref{flvslage}.
Models show the fluctuations getting brighter between 1 and 10 Myr, 
and progressively fainter after that. 
For $\barM(H)$ and especially for 
$\barM(K_s)$, after 10$^8$ years models with higher 
metallicity are brighter at a
fixed age (while for $\barM(J)$, at ages older than 1 Gyr,
the models are quite insensitive to metallicity).
This is mostly an effect of what we have
discussed earlier: more metal rich populations are redder and,
therefore, brighter in the near--IR. 
For example, the $K$-band magnitude of the tip of the 
RGB (TRGB) of Milky Way
clusters rises monotonically with metallicity, and is roughly 
one magnitude brighter at [Fe/H] $\sim$ -0.2 than at [Fe/H] $\sim$ -2.2
\citep{ferra00}.

The data follow the trend delineated by the
models, except that 
the behavior of the data between 1 and 50 Myr is less conspicuous
than the theoretical one,
owing mainly to the fact that the point 
for the SWB I supercluster is slightly fainter
than foreseen by the models.  
The brightness decline at the oldest ages is intensified by the
fact that the two oldest superclusters are of progressively 
lower metallicities.

The extreme and short--lived brightening of the fluctuation magnitude 
between 10$^6$ and 10$^7$ years is powered by red supergiants, 
while fluctuations of intermediate--age 
populations (500 Myr to 1 Gyr old) are fueled 
by intermediate mass stars in the AGB, which are 
brighter and redder then than the TRGB \citep[e.g.][]{frog80}.
For older populations, the fluctuations are driven by 
low mass stars in the RGB and AGB; 
the AGB is slightly bluer but not brighter than
the RGB at those stages. 
The gradual dimming of the fluctuation magnitude with age 
after 1 Gyr is a result of the reduction  
of the RGB average brightness.
This goes down, even though the 
luminosity of the TRGB stays constant and the 
IMF is rising, 
because lower mass stars
evolve at slower rates \citep{iben67}, 
and therefore spend longer and longer times at lower 
luminosities in the RGB. We illustrate these points with 
Figure \ref{isoch23}. 
It shows the evolution of the    
TRGB $K_s$ magnitude, 
the average RGB $K_s$ magnitude,  
and the $K_s$ magnitude of the mode of the 
RGB luminosity function, for a population with Z = 0.0004.
All magnitudes are in an arbitrary scale, and they 
are plotted vs main--sequence turn--off (MSTO) mass and log (age).
Between 400 Myr and $\sim$ 1 Gyr all functions 
first decrease, then increase in brightness. But after 1.5 Gyr 
the TRGB luminosity stays roughly constant, fueled by 
core helium ignition, while the average
RGB brightness decreases steadily. As a consequence
too of the onset of core degeneracy ---which means that energy is now 
furnished by H-shell burning only---, the brightness of
the mode of the RGB luminosity function plummets at 1 Gyr,
and subsequently dwindles with the progressively smaller MSTO.

\subsection{Fluctuation colors vs age} \label{fluctcol}

We display next plots of fluctuation color vs the logarithm of the age
(Fig.\ \ref{flcolvslage}).\footnote{The models with age = 10$^6$ yr are 
too blue and fall off the scale in ($\barJ\ - \barH$) and
($\barJ\ - \barKs$). On the other hand, the model with 
5 $\times$ 10$^8$ yr and Z = 0.008 is too red in ($\barJ\ - \barH$).}  
First focusing on the models in the top panels, we see that, 
like those for integrated colors, those with higher metallicities
are redder. 
However, rather than just reddening with age, they get 
redder until they reach a maximum, at 
$\sim$ 500 Myr; 
afterwards, those with   
metallicities of Z = 0.008 and higher become slightly bluer as the 
RGB takes over, 
but then roughly keep a constant color,
while the models with Z = 0.004 and lower turn bluer  
progressively thereafter. 
The reddest point at each metallicity is due to the cool, bright, AGB populated
by intermediate--mass stars in intermediate--age systems.

The behavior of the fluctuation color with age after $\sim$ 1 Gyr, 
for different metallicities, is a result of several features of the
RGB morphology: (1) we have already mentioned here that, for a single age,
the RGB temperature is lower for higher metallicities; (2)
for a single metallicity, the temperature of the RGB diminishes with 
age \citep{iben67,vand85}; (3) the amount 
of this difference in temperature between populations of different  
ages is a function of metallicity 
and in general is larger for higher Z \citep{vand85,rood97};
(4) the RGB is sloped, in the sense that more luminous stars are 
cooler than fainter ones, and the difference in temperature between
bright and faint stars increases with metallicity 
\citep[e.g.,][]{kuch95,ferra00}.\footnote{
Most of these characteristics are related to H$^-$ opacity, which in
turn is connected to metallicity, and 
to the dependence of ionization fraction on both 
atmospheric temperature and
density \citep{renz03}.
The RGB temperature diminishes with age because, since core
mass in this evolutionary phase is roughly independent from 
stellar main--sequence mass, 
a lower MSTO mass means a less massive  
atmosphere with less pressure which, consequently, expands and
cools more. The size of the RGB temperature difference with age depends 
on opacity. Initially, more metals will produce more 
continuum opacity, and hence more atmospheric expansion  
and lower temperature; a lower density will further 
increase the ionization fraction and the opacity. However, a point will be
reached when the low temperature will cause the 
ionization fraction and the opacity to diminish.
The RGB slope can be explained along similar lines: 
the ascent upwards the RGB is caused by 
atmospheric expansion. This expansion 
increases the stellar luminosity, but also decreases the density
of the atmosphere. With lower density, the ionization fraction 
and the opacity increase, with the subsequent diminution in temperature. 
The effect is stronger, and therefore the slope is larger, 
for higher metallicity.
}

Figure \ref{isoch8} shows how these features interplay with the 
fact ---discussed above in \S \ref{flucmag}--- that the RGB 
has a fainter average luminosity as a population ages,  
in order to determine the evolution of RGB color (and fluctuation color) 
with time, at several metallicities.
In this figure, we plot the average $K_s$ magnitude (arbitrary 
units) vs the average ($J - K_s$) integrated color of the 
RGB plus AGB, for populations with 
Z=0.004, Z=0.008, and Z=0.05.
The color is followed between 400 Myr 
(top of lines) and 20 Gyr (bottom of lines). 
In the case of the models we have used,
the maximum temperature difference of the RGB with age occurs at 
Z = 0.008. 
We see that the RGB plus AGB color of the 
Z = 0.004 model grows  
significantly bluer after 400 Myr and, despite reddening
later, is still bluer 
at 20 Gyr than at 400 Myr;  
the color reddens almost constantly 
for Z = 0.008; and it stays barely constant with time for Z = 0.05.
Hence, for Z = 0.004, the contribution from fainter,
hotter stars wins over the difference in temperature of the RGB
due to age, and over the smaller slope; 
the difference in temperature with age is dominant, even with a
steeper slope, 
for Z = 0.008; and all three effects compensate each other 
to yield a roughly constant temperature as the 
population evolves for Z=0.05. 
The progression of the RGB plus 
AGB integrated color seen in this graph is mirrored 
(albeit in an exaggerated fashion) 
by the model fluctuation colors in Figure \ref{flcolvslage}.

The behavior of our mixed metallicity star cluster data, on the other hand,
is much smoother than the models. 
This is not surprising, given the widths of our observed 
RGBs (Fig.\ \ref{cms}). 
The agreement between data and models 
is closer for (\barH\ - \barKs), 
but in all three fluctuation colors the data do not show a 
maximum. Instead, we can almost distinguish two
groups: the superclusters with log(age/yr) $<$ 8,
where massive main--sequence stars and red supergiants power the
SBFs, and those with
log(age/yr) $>$ 9, where fluctuations are produced by
low--mass stars in the RGB and AGB. There is an isolated point in these panels: 
the SWB III supercluster,
5$\times$10$^8$ years old, where red giants with intermediate mass
progenitors must be driving the SBFs; 
the coolest AGB stars that in theory dominate
intermediate--age populations in the near--IR do not
manifest themselves in the fluctuation colors of these data.
Finally, even though we do see the bluer color at 
older ages that can be explained by the progressively 
lower metallicities of superclusters VI and VII, the
data overall display  
a tendency to redden with age that can be fit
by a straight line. 

\noindent
Thus, we find, {\it for the  
superclusters only}:

\begin{equation}
(\barJ\ - \barH) = (0.10\pm0.02){\rm log (age/yr)} + (0.29\pm0.19) 
\end{equation}

\begin{equation}
(\barH\ - \barKs) = (0.21\pm0.03){\rm log (age/yr)} - (1.29\pm0.22)
\end{equation}

\begin{equation}
(\barJ\ - \barKs) = (0.30\pm0.04){\rm log (age/yr)} - (0.97\pm0.36). 
\end{equation}

\noindent
About one third of the galaxies, on the other hand, 
fall at a redder (\barH\ - \barKs) color than the 
superclusters and all of the models. If, this notwithstanding, 
we include them in the linear fit, we get 

\begin{equation}
(\barH\ - \barKs) = (0.20\pm0.02){\rm log (age/yr)} - (1.25\pm0.16).
\end{equation}

\noindent
Table \ref{tabflcol} lists the coefficients of these fits. 
The reduced chi--square ($\widetilde{\chi}^2$) and the rms 
of the points (in magnitudes) after the fits are also included.


\subsection{Trends with metallicity} \label{metals}

For completeness, we present plots of fluctuation magnitudes 
and colors vs metallicity. 
On average, populations with higher metallicities are
brighter and redder at the same age,
but the behavior followed by the models here
can be understood in more detail from
our exposition of RGB and AGB brightnesses
and colors in the subsections above. 

Fig.\ \ref{flvsz} displays the \barJ, \barKs, and 
\barH\ fluctuation magnitudes vs metallicity.
Here, the match between models and data is in general good.
The data points for the MC superclusters, in the low metallicity 
(left) region of the plots, show once more a general trend with age, where 
younger populations are brighter than older ones ---although 
in particular supercluster Pre-SWB is much brighter than the
models, and superclusters classes V, VI, and VII are somewhat brighter
than the models. The Fornax Cluster galaxies, on the other hand, 
clump up in the high metallicity (right) regions of the middle and 
bottom panels, where models 
2 Gyr and older ``funnel'' and loose the ability to 
make fine distinctions in age.

In Fig.\ \ref{flcolvsz}, we plot the (\barH\ - \barKs) 
fluctuation color vs metallicity. In order to facilitate the
comparison with the models, the top panels show the MC clusters 
only, while the bottom ones include the Fornax galaxies also.
The Pre-SWB supercluster is redder than the models, while 
classes III, VI and VII are somewhat bluer, but data and
models mirror each other in the tendency for objects 
up to 5 Gyr to redden with age; the two oldest superclusters 
then become bluer, owing both to age (see \S \ref{fluctcol}) 
and their lower metallicities. 
On the other hand, a few of the Fornax galaxies seem significantly
redder than the reddest models, and indeed the galaxy average,
although matching the models between 500 Myr and 2 Gyr, 
is only marginally consistent with those between 
5 and 17 Gyr old. (We remind the reader
that the error bars of the galaxy fluctuation color 
have been calculated using
information from the literature and assuming 
a correlation coefficient of -0.9 between any two given 
passbands, while errors for SBFs of the MC superclusters have been 
derived directly from the data.)  

\section{Conclusions and Future Work} \label{disc}


%



This study has shown that in MC star clusters, most of which
have roughly the same relatively low metallicity,
near--IR fluctuation magnitudes and colors are
driven by age. 

Our result is not unexpected. In their classical study, 
\citet{sear80} demonstrated that the properties of the MC
clusters' integrated light {\it in the optical wavelengths}
are determined by their red giants. They also inferred  
that the sequence from class I to class III is one of age, and 
insensitive to abundance.
Regarding the sequence of the older clusters, classes IV through VII,
\citet{sear80} posited that it was both sensitive to 
increasing age {\it and} decreasing metallicity. 
Given that (a) both near--IR wavelengths and 
SBFs are more sensitive than integrated optical 
light to the red giant stars in these clusters,  
and (b) the star cluster metallicity stays 
nearly the same for classes Pre through V, 
it is not surprising that we have also found a sequence of
age, slightly modulated by abundance in the case of the two oldest 
SWB classes.

It is true that 
the MC star clusters are mostly either too young 
(7 out of the 8 MC superclusters are younger than
most of the Fornax galaxies)
and/or too metal--poor to
be relevant to the galaxies.
However they seem to 
outline a trend with age that includes the
galaxies, as is shown most clearly by Figure \ref{flvslage}.
For this reason, even if star cluster populations might be of 
limited direct value for the modelling of old ellipticals and
spheroids, they are important for the calibration of stellar
population synthesis models.

Regarding the agreement between the data and the 
models in their present state, we find that it is 
very good qualitatively, but that it could be improved in
the details. For example, 
in principle, we could have read off the metallicity 
of the superclusters from Figures \ref{flvslage} and \ref{flcolvslage}, 
given their ages.
However, the metallicities we would have inferred 
are not consistent in all cases with the ones 
that correspond to their SWB class. Also, the models, and 
in particular those with the highest metallicity, cannot reproduce 
the very red fluctuation colors exhibited by a few of the 
Fornax galaxies. Conversely, models predict redder fluctuation colors 
than those of intermediate--age clusters, and they underestimate the
contribution of bright stars to the integrated luminosity in 
these same clusters.   
Arguably, the models work best for old, metal poor populations.
This is probably not a coincidence, but is due to the fact
that models have tried to match their features for the longest time.
Moreover, old populations evolve more slowly. On the 
MC supercluster data side, the oldest clusters are also the 
most massive, and therefore have the smallest stochastic
errors. The rapidly evolving young populations are harder to
match, as are the intermediate--age ones, with their 
poorly, albeit increasingly better, understood asymptotic
giant branches.

We plan to continue this work in various directions, e.g.,
improve the calibration of the models with the SBF data, 
compare to other models, and
compute SBFs of the highest metallicity clusters in our Galaxy
and in M~31. We will also investigate the 
relationship between \barM$_{K_s}$ and ($V - I$) color in the 
MC star clusters. \citet{liu02} discovered  
a linear dependence of \barM$_{K_s}$ with ($V - I$) in a sample of 
26 ellipticals, S0s, and spiral bulges, which might
be tracing late bursts of star formation in these systems; 
a study of the MC star clusters, which probe a range twice as large 
in \barM$_{K_s}$ and three times larger in color, 
is likely to throw light into the origin of this trend. 
 

\acknowledgments

This research has made use of the VizieR Service 
and the SIMBAD database at
the Centre de Donn\'ees Astronomiques de Strasbourg, as well as NASA's
Astrophysics Data System Abstract Service.
R.A.G.\ acknowledges 
support from two Mexican organizations, respectively, the 
National Researcher System
(SNI) and the National Council of Science and 
Technology, CONACyT (grant 36042-E); she also 
thanks A.\ Watson for suggesting the use of the 
2MASS database, P.\ D'Alessio for her help with 
plotting macros, M.\ Cervi\~no for  
lively discussions regarding errors, 
S.\ Van Dyk for dissipating doubts about
the 2MASS survey, 
and L.\ Loinard for his willingness 
to discuss every aspect of this work.  
M.C.L.\ is grateful to research support from NASA grant
HST-HF-01152.01-A. GBA acknowledges support from 
the Venezuelan Ministerio de Ciencia y 
Tecnolog\'{\i}a and FONACIT. We express our appreciation to the 
anonymous referee, whose pertinent and helpful suggestions
allowed us to improve the robustness of our results. 
It is our pleasure to 
thank James R.\ Graham for inspiring this work. 

\appendix

\section{Error propagation} \label{errprop}

Following \citet{buzz89} and \citet{cerv02}, one can 
estimate the relative error, in the Poissonian limit, for any
synthesized quantity $A$ which is the sum of the contributions 
from individual stars (or populations); i.e.,  
$A=\sum w_i a_i$, 
where $a_i$ is the contribution of the $i^{th}$ star (or stellar type), 
and $w_i = N_i / M_{tot}$ is the mean value of the number of stars of
mass $m_i$, normalized to the total mass of the cluster 
$M_{tot} = \sum m_i$ (assumed to be a constant).
$w_i$ is treated as a random variable, but $a_i$ is considered
a fixed quantity; in this case, 

\begin{equation}
\sigma^2_A = \sum a_i^2 \sigma^2_{w_i} .
\end{equation}

\noindent Now, 

\begin{equation}
\sigma^2_{w_i} = \frac{N_i}{M_{tot}^2} = \frac{w_i}{M_{tot}},
\end{equation}

\noindent so that the relative error is

\begin{equation}
\frac{\sigma_A}{A} =\frac{\left( \frac{1}{M_{tot}} \sum w_i a_i^2 \right)^{1/2}}{\sum w_i a_i}.
\label{relerr}
\end{equation}

Notice that the importance of the 
stochastic fluctuations 
in the number of contributing stars goes down as the total mass
of the cluster increases. 

\subsection{Error of an integrated monochromatic luminosity.} \label{eml}

In this case, Eq.\ \ref{relerr} applies directly. Suppose that $J$ is the
integrated luminosity in
the $J$-band and $j_i$ is the contribution from the $i^{th}$ stellar type; then

\begin{equation}
\frac{\sigma_J}{J} =\frac{\left( \frac{1}{M_{tot}} \sum w_i j_i^2 \right)^{1/2}}{\sum w_i j_i}.
\end{equation}

The error in magnitudes is the relative error $\times$ 2.5 $\times {\rm log}_{10}(e)$, or

\begin{equation}
1.0857 \times \frac{\sigma_J}{J}\ \ {\rm mag}
\end{equation}

\subsection{Error of an integrated color.} \label{eic}

We begin by expressing the color as the ratio of two luminosities, or

\begin{equation}
c = \frac{\sum w_i a_i}{\sum w_i b_i} \equiv \frac {u}{v};
\end{equation}

\noindent $a_i$ and $b_i$ are, respectively, the contributions of stars of type $i$ in each
wavelength. In this case, the fluxes in the two bands {\it are} correlated, but
we assume that $w_i$ is independent from both $a_i$ and $b_i$.
We will apply the following equation for the relative error of a random variable 
$z= x / y$:

\begin{equation}
\frac{\sigma^2_z}{z^2} \simeq \frac{\sigma^2_x}{x^2} + \frac{\sigma^2_y}{y^2} - 2\frac{{\rm cov}(x,y)}{xy}. 
\end{equation}

So,

\begin{equation}
\sigma^2_u =  \frac{1}{M_{tot}} \sum w_i a_i^2
\end{equation}

\begin{equation}
\sigma^2_v =  \frac{1}{M_{tot}} \sum w_i b_i^2
\end{equation}

\begin{equation}
{\rm cov}(u,v) = \frac{1}{M_{tot}} \sum w_i a_i b_i.  
\end{equation}

Hence, the relative error square is


\begin{equation}
\frac{\sigma^2_c}{c^2} \simeq \frac{\frac{1}{M_{tot}} \sum w_i a_i^2}{\left( \sum w_i a_i \right)^2} + 
\frac{ \frac{1}{M_{tot}} \sum w_i b_i^2}{\left( \sum w_i b_i \right)^2} - 
\frac{ \frac{2}{M_{tot}} \sum w_i a_i b_i}{\sum w_i a_i \sum w_i b_i}, 
\end{equation}

\noindent
and the relative error in magnitudes is

\begin{equation}
1.0857 \times \left( \frac{\sigma^2_c}{c^2} \right)^{1/2}\ \ {\rm mag}
\end{equation}

\subsection{Error of a fluctuation luminosity.} \label{efl}

This is again the case of a ratio:

\begin{equation}
\barl = \frac{\sum w_i a_i^2}{\sum w_i a_i} \equiv \frac {u}{v};
\end{equation}

\noindent $a_i$ and $a_i^2$ are, of course, correlated. So,

\begin{equation}
\sigma^2_u = \frac{1}{M_{tot}} \sum w_i a_i^4 
\end{equation}

\begin{equation}
\sigma^2_v = \frac{1}{M_{tot}} \sum w_i a_i^2
\end{equation}

\begin{equation}
{\rm cov}(u,v) = \frac{1}{M_{tot}} \sum w_i a_i^3,  
\end{equation}

\noindent and


\begin{equation}
\frac{\sigma^2_{\footnotesize \barl}}{\footnotesize \barl^2} \simeq \frac{ \frac{1}{M_{tot}} \sum w_i a_i^4}{(\sum w_i a_i^2)^2} + 
\frac{ \frac{1}{M_{tot}} \sum w_i a_i^2}{(\sum w_i a_i)^2} - 
\frac{ \frac{2}{M_{tot}} \sum w_i a_i^3}{\sum w_i a_i^2 \sum w_i a_i};
\end{equation}

\noindent the relative error in magnitudes is

\begin{equation}
1.0857 \times \left( \frac{\sigma^2_{\footnotesize \barl}}{\footnotesize \barl^2} \right)^{1/2}\ \ {\rm mag}
\end{equation}

\subsection{Error of a fluctuation color.} \label{efc}

We express the fluctuation color as follows:

\begin{equation}
fc = \frac{\sum w_i a_i^2 / \sum w_i a_i}{\sum w_i b_i^2 / \sum w_i b_i} =
\frac{\sum w_i b_i}{\sum w_i a_i} \cdot \frac{\sum w_i a_i^2}{\sum w_i b_i^2}
\equiv \frac {u}{v} \cdot \frac{r}{s};
\end{equation}

\noindent
since $a_i$, $a_i^2$, $b_i$, and $b_i^2$ are all correlated,

\begin{eqnarray}
\frac{\sigma^2_{\footnotesize fc}}{\footnotesize (fc)^2} \simeq
\frac{\sigma^2_u}{u^2} + \frac{\sigma^2_r}{r^2} + 2 \frac{{\rm cov}(u,r)}{ur} +
\frac{\sigma^2_v}{v^2} + \frac{\sigma^2_s}{s^2} + 2 \frac{{\rm cov}(v,s)}{vs} \nonumber \\ 
- 2 \frac{{\rm cov}(u,v)}{uv} - 2 \frac{{\rm cov}(r,v)}{rv}  
- 2 \frac{{\rm cov}(u,s)}{us} - 2 \frac{{\rm cov}(r,s)}{rs}  
\end{eqnarray}

\begin{equation}
\sigma^2_u = \frac{1}{M_{tot}} \sum w_i b_i^2
\end{equation}

\begin{equation}
\sigma^2_r = \frac{1}{M_{tot}} \sum w_i a_i^4
\end{equation}

\begin{equation}
\sigma^2_v = \frac{1}{M_{tot}} \sum w_i a_i^2
\end{equation}

\begin{equation}
\sigma^2_s = \frac{1}{M_{tot}} \sum w_i b_i^4
\end{equation}

\begin{equation}
{\rm cov}(u,r) = \frac{1}{M_{tot}} \sum w_i a_i^2 b_i
\end{equation}

\begin{equation}
{\rm cov}(v,s) = \frac{1}{M_{tot}} \sum w_i a_i b_i^2
\end{equation}

\begin{equation}
{\rm cov}(u,v) = \frac{1}{M_{tot}} \sum w_i a_i b_i
\end{equation}

\begin{equation}
{\rm cov}(r,v) = \frac{1}{M_{tot}} \sum w_i a_i^3
\end{equation}

\begin{equation}
{\rm cov}(u,s) = \frac{1}{M_{tot}} \sum w_i b_i^3
\end{equation}

\begin{equation}
{\rm cov}(r,s) = \frac{1}{M_{tot}} \sum w_i a_i^2 b_i^2
\end{equation}

Therefore,

\begin{eqnarray}
M_{tot} \cdot \frac{\sigma^2_{\footnotesize fc}}{\footnotesize (fc)^2} \simeq
\frac{\sum w_i b_i^2}{\left(\sum w_i b_i \right)^2} 
+ \frac{\sum w_i a_i^4}{\left(\sum w_i a_i^2 \right)^2}
+ 2 \frac{\sum w_i a_i^2 b_i}{\sum w_i a_i^2 \sum w_i b_i} \nonumber \\
+ \frac{\sum w_i a_i^2}{\left(\sum w_i a_i \right)^2} 
+ \frac{\sum w_i b_i^4}{\left(\sum w_i b_i^2 \right)^2}
+ 2 \frac{\sum w_i a_i b_i^2}{\sum w_i a_i \sum w_i b_i^2} \nonumber \\
- 2 \frac{\sum w_i a_i b_i}{\sum w_i a_i \sum w_i b_i} 
- 2 \frac{\sum w_i a_i^3}{\sum w_i a_i \sum w_i a_i^2} \nonumber \\ 
- 2 \frac{\sum w_i b_i^3}{\sum w_i b_i \sum w_i b_i^2} 
- 2 \frac{\sum w_i a_i^2 b_i^2}{\sum w_i a_i^2 \sum w_i b_i^2}
\end{eqnarray}

\noindent
and the relative error in magnitudes is

\begin{equation}
1.0857 \times \left( \frac{\sigma^2_{\footnotesize fc}}{\footnotesize f^2c^2} \right)^{1/2}\ \ {\rm mag}
\end{equation}

\clearpage

\figcaption{
Comparison between the stellar transformation from 
\HST\ NICMOS Camera 2 $F160W$ to CIT/CTIO $H$ (\citet{step00}, {\it dotted 
lines}),  
and the SBF transformations derived from stellar population synthesis
models. {\it Top panels}: solar metallicity; {\it bottom panels}: Z = 0.05.
{\it Left column}: \barH\ - \barF vs ($J - K$); 
{\it middle column}: vs ($J - K$)$_{bright}$ of stars 
brighter than $M_{K}$ = -4.5; {\it right column}: 
vs (\barJ\ - \barK). The model transformations are closest
to the stellar transformation when using the color of
the brightest stars; ($J - K_s$)$_{bright}$ is about 
0.48 mag redder than the color of the whole population for 
0.85 $\geq$ ($J - K_s$) $\leq$ 0.95.
\label{ftransf}}

\figcaption{MC superclusters. Greyscale versions of $J$, $H$, and 
$K_s$ color mosaics built with 2MASS data. N is up and E is to 
the left; images are 3\farcm3 on the side. 
\label{mosaics}}

\figcaption{Color--magnitude diagrams of MC superclusters.
Stars 
within 60 arcsec from the center (at the distance of the LMC).
Average photometric errors are 0.04 mag               
in brightness and
0.02 mag in color for sources with
$K_s \leq$ 13; 0.06 and 0.03 mag 
for stars with 13 $< K_s \leq$ 14; and 0.13 and 0.07 mag
(about the size of the dots) for sources with 14 $< K_s \leq$ 15.
\label{cms}}

\figcaption{Comparison between contributions from all stars in the 
isochrone and only from stars brighter than $M_{K_s}$ = -4.5
(or $K_s$ = 14 at the LMC). 
{\it Top left panel}: Z = 0.0004; {\it top right panel}: 
Z=0.004; {\it bottom left panel}: Z=0.008; {\it bottom right panel}:
Z=0.05. Within each panel, 
difference in magnitudes 
at $J$ ({\it solid line}),
$H$ ({\it dotted line}), $K_s$ ({\it dashed line}), 
\barJ\ ({\it long--dashed line}), \barH\ ({\it dotted--dashed line}),
and \barKs\ ({\it dotted--long--dashed line}). 
\label{convtest}}

\figcaption{Systematic errors in MC supercluster
parameters. {\it Top}: $H$-band SBF measurements vs log (age) of
all eight MC superclusters; 
{\it middle}: ($\barH\ - \barKs$) colors; {\it bottom}: 
ratio of observed to model contributions of 
stars brighter than $M_{K_s}$ = -4.5 to integrated light at 
$H$.
{\it Left panels:} Point sources with bad photometry have
been eliminated. {\it Right panels:} All point sources from the
PSC have been included. Dots of different colors represent
values obtained from different radial ranges. Error bars are 
Poisson. Lines that represent models with different metallicities 
(see Figures \ref{flvslage} and \ref{flcolvslage}) are included 
in top and middle panels. 
\label{figsyst}}

\figcaption{{\it Left}: comparison of integrated colors
vs log(age) with stellar population synthesis models.
{\it Top}: ($J - H$); {\it middle}: ($H - K_s$); 
{\it bottom}: ($J - K_s$). Filled symbols are MC clusters
from this work, and open ones are Fornax Cluster galaxies
whose colors we derived from the 2MASS Extended Source 
Catalog. Models of a fixed metallicity have the same symbol, with increasing
symbol size representing increasing age. Models have
Z = 0.0004, 0.004, 0.008, and 0.05; 
they are 1, 10, 50 and 500 Myr, 2, 5, 8, and 17 Gyr old. {\it Right}: 
the same data points are plotted with
error bars. The values in parentheses are the number of objects
in each group.
\label{figcol}}

\figcaption{{\it Left}: comparison of $J$-band 
({\it top}), $H$-band ({\it middle}), 
and $K_s$-band ({\it bottom}) SBF measurements
vs age with stellar population synthesis models.
Models and symbols are the same as in Figure \ref{figcol}. 
{\it Right}: the same data points are plotted with
error bars. The values in parentheses are the number of objects
in each group with SBF measurements.
\label{flvslage}}

\figcaption{$K_s$ magnitudes in an arbitrary scale, 
vs main sequence turn--off mass and log (age), 
for a population with Z = 0.0004.
{\it Solid line}: average 
of the RGB; {\it short--dashed line}: TRGB; 
{\it dotted line}: mode of the RGB luminosity function. 
\label{isoch23}}

\figcaption{{\it Top}: Comparison of SBF colors
vs log (age) with stellar population synthesis models.
Models and symbols are the same as in previous figures.
{\it Left}: (\barJ\ - \barH); {\it middle}: (\barH\ - \barKs);
{\it right}: (\barJ\ - \barKs).
{\it Bottom}: the same data points are plotted with
error bars. The values in parentheses are the number of objects
in each group with SBF measurements.
Dashed lines are best fits {\it to MC star clusters};
dotted line is best fit of (\barH\ - \barKs) vs log (age) 
for MC superclusters {\it and} Fornax galaxies (see text).
\label{flcolvslage}}

\figcaption{Evolution of the RGB plus AGB average 
$K_s$ magnitude (arbitrary scale) vs average color, 
between 400 Myr (top of lines) and 20 Gyr (bottom
of lines),
for populations with 
different metallicities. {\it Solid line}: Z = 0.004; 
{\it dotted line}: Z = 0.008; {\it short--dashed line}:
Z = 0.05. Tick marks at 0.4, 1, 4, 16, and 20 Gyr.
\label{isoch8}}

\figcaption{{\it Left}: comparison of $J$-band ({\it top}), 
$H$-band ({\it middle}), 
and $K_s$-band ({\it bottom}) SBF measurements
vs metallicity with stellar population synthesis models.
Models and symbols are the same as in previous figures.
Lines connect models with the
same age; models are 50 and 500 Myr, 2, 5, 8, and 17 Gyr old.
{\it Middle}: the same data points are plotted with
error bars. The values in parentheses are the number of objects
in each group with SBF measurements. {\it Right}: the scale has been changed
to show loci of 1 and 10 Myr old models; insets show location of
plots in the left and middle.
\label{flvsz}}

\figcaption{{\it Top left}: comparison of (\barH\ - \barKs) SBF
color vs metallicity of MC superclusters with stellar 
population synthesis models. 
Models, symbols, and lines are the same as in figure
\ref{flvsz}.
{\it Bottom left}:
comparison of (\barH\ - \barKs) SBF
color vs metallicity of MC superclusters {\it and} 
Fornax Cluster galaxies with stellar
population synthesis models; the scale has been changed 
relative to the top left panel to allow for the inclusion
of the galaxies.
Models and symbols are the same as in previous figures. 
The green error bars represent the galaxy average, which is only
marginally consistent with the models between 5 and
17 Gyr old.
{\it Middle}: the same data points are plotted with
error bars. The values in parentheses are the number of objects.
{\it Right}: the scale has been changed anew
to show loci of 1 and 10 Myr old models; insets show location of
plots in the left and middle.
\label{flcolvsz}}

\clearpage
\begin{landscape}
\begin{deluxetable}{lccclccccc}
\tablewidth{8.45in}
\tablecaption{Cluster Data}
\tablehead{
\colhead{Supercluster} &
\colhead{Age (yr)} &
\colhead{Z} &
\colhead{Region} &
\colhead{Name} &
\colhead{$N_*$} &
\colhead{$s$-parameter} &
\colhead{Cloud} &
\colhead{$E(B-V)$} &
\colhead{SWB-class}\\[-0.3cm]
}

\startdata

Pre-SWB\dotfill& 2.4$\times$10$^6$& 0.01 & 0\arcsec --- 60\arcsec & IC~2128 &  \ 9 &  \ 1&   LMC & & \\ 
&&&&NGC~1748 &    16 &  \ 1&   LMC & & \\ 
&&&&NGC~1743&   19 &   \ 2&    LMC & & \\
&&&&L~107 &  \ 3 &   \ 3 &  SMC             & & \\
&&&&NGC~1714&  \ 6 &   \ 3&    LMC& & \\
&&&&NGC~1727$^{\rm a}$ &   22 & \ 4 & LMC & & \\
&&&&NGC~1910$^{\rm a}$ &  34 & \ 4 & LMC & & \\
&&&&NGC~1936 (IC~2127)$^{\rm a}$ &  \ 8 & \ 4 & LMC & & \\
&&&&L~84$^{\rm a}$ &  25 & \ 5 & SMC & & \\
&&&&NGC~602&  \ 6 &    \ 6&    SMC& & \\
&&&&NGC~2001$^{\rm a}$ &  20 & \ 6 & LMC & & \\
&&&&NGC~1833$^{\rm a}$ &  17 & \ 7 & LMC & & \\
&&&&NGC~2027&  12 &   \ 7&    LMC & & \\
&&&&SL~362$^{\rm a}$ &  39 & \ 7 & LMC & & \\
&&&&NGC~2014$^{\rm b}$ &  10 &  \ 8&  LMC& & \\
&&&&NGC~346   &  15 & \ 8&    SMC& & \\
&&&&HS~314 &   31 & 10& LMC& & \\
&&&&NGC~2074$^{\rm b}$&  20 &  10&  LMC& & \\
&&&&SL~360$^{\rm a}$ & 33 & 10 & LMC & & \\
&&&&NGC~1984&  25 &  11&   LMC&      0.15 & \\        
&&&&NGC~2018&  20 &  11&    LMC& & \\
&&&&NGC~1873$^{\rm a}$ &  10 & 12 & LMC & & \\
&&&&L~70  &  \ 8 &  13  &  SMC & & \\
&&&&NGC~2006&  \ 9 &  13&    LMC   &  & \\                  
&&&&NGC~1983 &  28 & 13&   LMC & & \\
&&&&NGC~2011 &  \ 4 & 13&   LMC    &  0.08 & \\     
&&&&SL~114&  24 &  13  &  LMC& & \\
&&&&L~74   &  10 &  14  &  SMC & & \\
SWB I\dotfill & 1$\times$10$^7$& 0.01& 0\arcsec --- 60\arcsec & NGC~2003& 
 \ 2 & 15&    LMC & & \\
&&&&L~51&   \ 7 &   15&    SMC& & \\
&&&&L~48&   \ 3 &   15&    SMC& & \\
&&&&NGC~1994&  22 &  15&   LMC&      0.14 & \\
&&&&NGC~2004&  15 &  15&   LMC&      0.06 &                I \\
&&&&SL~538&  10 & 15&    LMC & & \\
&&&&L~56&   15 &   16&   SMC&          &        II \\
&&&&NGC~290&  20 &   16&    SMC& & \\
&&&&NGC~1767&  25 &  16&    LMC&        0.15 & \\
&&&&NGC~1787&  20 &  16&    LMC& & \\
&&&&NGC~2009&  13 &  16&    LMC& & \\
&&&&NGC~2098&  \ 9 &  16&    LMC& & \\                       
&&&&L~45$^{\rm a}$& 15 & 17 & SMC & & \\
&&&&NGC~1766&  11 &  17&    LMC&  & \\
&&&&NGC~1772& 24 &    17&    LMC&  & \\
&&&&NGC~1805&  12 &  17&    LMC&         0.10  & \\        
&&&&NGC~2002& \ 5 &  17&    LMC& & \\ 
&&&&NGC~2100&  21 &  17&   LMC&      0.24     &            I \\
&&&&L~66&   15 &   18&    SMC& & \\
&&&&NGC~1810&  12 & 18&   LMC& & \\
&&&&NGC~1818&  11 &  18&   LMC&      0.10 &                I \\
&&&&NGC~330&  22 &   19&   SMC&           &           I \\
&&&&NGC~176&  \ 8 &   20&   SMC& & \\                  
&&&&NGC~299&  \ 9 &  20&   SMC&                   &          I \\
&&&&NGC~1704&  10 &  20&    LMC& & \\
&&&&NGC~1711&  \ 9 &  20&    LMC&        0.16             &   II \\
&&&&NGC~1860&  40 &  20&   LMC& & \\                        
&&&&NGC~376&  21 &   20&    SMC& & \\
&&&&SL~477&   12 & 20&    LMC& & \\
SWB II\dotfill & 5$\times$10$^7$& 0.01& 0\arcsec --- 60\arcsec & NGC~1869&  
12 &   21&    LMC & & \\
&&&&NGC~1698&  15 &   21&    LMC & & \\
&&&&NGC~1847&  32 &  21&   LMC & & \\                        
&&&&NGC~1850&  43 &  21&   LMC&      0.15 & \\              
&&&&NGC~1863&  37 &  21&   LMC & & \\
&&&&SL~106&  25 &  21&    LMC & & \\
&&&&IC~1612&  16 &  22&    SMC & & \\
&&&&L~39&     18 & 22&    SMC & & \\
&&&&NGC~220&  \ 9 &   22&   SMC  & &                           III \\
&&&&NGC~222&  \ 8 &   22&   SMC  & &                            II-III \\
&&&&NGC~1735&  14 &  22&    LMC & & \\
&&&&NGC~1793&  16 &  22&    LMC & & \\
&&&&NGC~1834&  42 &  22&   LMC & & \\
&&&&NGC~1855&  65 &  22&    LMC&        0.12 & \\
&&&&NGC~1928$^{\rm a}$ & 46 & 22 & LMC & & \\
&&&&NGC~2214$^{\rm a}$ & 14 & 22 & LMC & 0.10 & II \\
&&&&IC~1624&   21 & 23&    SMC & & \\
&&&&NGC~1774&  15 &  23&    LMC &        0.10      &           II \\
&&&&NGC~1782& 21 &  23&    LMC & & \\
&&&&NGC~1804&  29 &  23&    LMC & & \\
&&&&NGC~1903$^{\rm a}$ & 34 & 23 & LMC & & \\
&&&&NGC~2164&  25 &  23&   LMC &     0.10    &                   III \\
&&&&IC~1655$^{\rm c}$& \ 5 & 24&    SMC & & \\
&&&&NGC~231&  12 &   24&    SMC & & \\
&&&&NGC~242&  13 &   24&   SMC  & &                           II \\
&&&&NGC~422&  \ 6 &   24&   SMC & & \\                  
&&&&NGC~1732&  12 &  24&    LMC & & \\
&&&&NGC~1755&  16 &  24&    LMC &       0.12 &                II-III \\
&&&&NGC~1854&  64 &  24&   LMC  &    0.13   &                    II \\
&&&&NGC~1870&  30 &  24&   LMC & & \\
&&&&NGC~1913$^{\rm a}$ & 30 & 24 & LMC & & \\
&&&&NGC~1951&  17 &  24&    LMC &       0.10 &  \\
&&&&NGC~1986&  52 & 24&    LMC &       0.18 &                    II \\
&&&&NGC~2118&  12 &  24&    LMC & & \\
&&&&SL~56&    10 & 24&    LMC & & \\
SWB III\dotfill & 5$\times$10$^8$& 0.01& 0\arcsec --- 60\arcsec & IC~1660& 
\ 5 &  25&    SMC & & \\
&&&&L~44&  30 &   25&    SMC & & \\
&&&&L~63&  16 &   25&    SMC & & \\
&&&&NGC~256&  14 &  25&   SMC    & &                          II \\
&&&&NGC~458&  23 &  25&   SMC    & &                         III \\
&&&& NGC~1828& 34 &   25&    LMC & & \\ 
&&&&NGC~1844& 20 &  25&   LMC & & \\                         
&&&&NGC~1943$^{\rm a}$ & 41 & 25 & LMC & 0.18 & \\
&&&&NGC~2000& 20 &  25&    LMC & & \\
&&&&NGC~2041& 15 &  25&    LMC  &      0.05  &              III \\
&&&&NGC~2157& 17 &  25&   LMC &     0.10 & \\             
&&&&NGC~2159& 17 &  25&   LMC &     0.10 & \\                
&&&&NGC~2172& 10 &  25&   LMC &     0.10 & \\              
&&&&SL~539& 27 &  25&    LMC & & \\
&&&&IC~1611$^{\rm a}$ & 17 & 26 & SMC & & \\
&&&&NGC~265& 33 &   26&   SMC    & &                        III \\
&&&&NGC~2058& 57 &   26&   LMC   &   0.18 &                     III \\
&&&&NGC~2065& 65 &   26&   LMC   &   0.18 &                     III \\
&&&&NGC~2136& 31 &   26&   LMC   &   0.10 &                     III \\
&&&&NGC~2156&  11 &  26&   LMC   &   0.10 & \\              
&&&&L~40$^{\rm a}$ & 20 & 27 & SMC & & \\
&&&&NGC~1866 &  37 & 27&   LMC   &   0.10 &                      III \\
&&&&NGC~2025& 26 &  27&    LMC & & \\
&&&&NGC~2031&  38 & 27&    LMC & & \\                          
&&&&L~114&   \ 4 &  28&    SMC & & \\
&&&&NGC~1775$^{\rm a}$ & 18 & 28 & LMC & & \\
&&&&NGC~1885 &  46 & 28&    LMC & & \\
&&&&NGC~1895$^{\rm a}$ & 10 & 28 & LMC & & \\
&&&&NGC~2134&   42 & 28&   LMC &     0.10    &                   IV \\
&&&&NGC~269&  15 &  29&   SMC     &     &                 III-IV \\
&&&&NGC~1830& 35 &   29&    LMC & & \\
&&&&NGC~1953& 25 &  29&    LMC    &    0.12 & \\
SWB IV\dotfill& 1$\times$10$^9$& 0.01& 0\arcsec --- 60\arcsec & L~53& 
 16 &   30&    SMC& & \\
&&&&NGC~294 (L~47)&   24 &  30&   SMC& & \\                          
&&&& NGC~1801& 52 & 30 & LMC& & \\ 
&&&&NGC~1856$^{\rm a}$ & 58 & 30 & LMC& 0.24 & IV \\
&&&&NGC~1872&  51 & 30&  LMC&   0.13&                III-IV \\
&&&&NGC~1831$^{\rm a}$ & \ 6 & 31 & LMC& 0.10 & V \\
&&&&NGC~2056&  41 &  31&    LMC& & \\
&&&&L~26$^{\rm a}$ & \ 4 & 32 & SMC & & \\
&&&&NGC~1756&  29 &  32&    LMC& & \\
&&&&NGC~1849&  19 &  32&    LMC& & \\
&&&&NGC~2107&  46 &  32&    LMC &       0.19 &                  IV \\
&&&&SL~562&  23 &  32&    LMC& & \\
&&&&NGC~1868&  \ 5 &  33&   LMC &     0.07   & \\         
&&&&NGC~2249&  11 &  34&   LMC& & \\                       
&&&&NGC~1987&  27 &  35&    LMC&        0.12&                       IV\\
&&&&NGC~2209&  10 & 35&   LMC&      0.07  &                     III-IV\\
&&&&NGC~2108&  16 &  36&    LMC&        0.18& \\
&&&&SL~663$^{\rm a,b}$ &  15 & 36 & LMC & & \\
\
\
\tablebreak
SWB V\dotfill& 3$\times$10$^9$& 0.008 & 0\arcsec --- 60\arcsec & IC~2146 & 
19 &   37 &   LMC & & \\  
&&&&NGC~152  &  21 & 37 & SMC   &        &        IV      \\
&&&&NGC~411  &  15 & 37 & SMC   &        &         V-VI   \\
&&&&NGC~1644$^{\rm a}$ & 10 & 37 & LMC & & \\
&&&&NGC~1783 &  33 & 37 & LMC   &   0.10 &         V      \\
&&&&NGC~2231 &  22 & 37 & LMC  &   0.10 &         V      \\
&&&&SL~363$^{\rm a}$ & 41 & 37 & LMC & & \\
&&&&NGC~419  &  25 & 38 & SMC   &        &         V       \\
&&&&NGC~1777$^{\rm b}$&  10 & 38 & LMC  &        &                 \\
&&&&SL~556$^{\rm a,b}$ & 18 & 38 & LMC & & \\
&&&&NGC~1651$^{\rm a}$ & 20 & 39 & LMC & & \\
&&&&NGC~1917$^{\rm a}$ & 36 & 39 & LMC & & \\
&&&&NGC~2154 &  29 & 39 & LMC   &    0.10&         V       \\
&&&&NGC~2162 &  \ 8 & 39 & LMC  &   0.07 &         V       \\
&&&&NGC~2213 &  16 & 39 & LMC  &   0.10 &         V-VI    \\
&&&&NGC~1806 &  50 & 40 & LMC   &    0.12&         V       \\
&&&&NGC~1846$^{\rm a}$ & 38 & 40 & LMC & 0.10 & V \\
&&&&NGC~2193$^{\rm b}$&  16 & 40 & LMC  &        &                  \\
&&&&SL~855$^{\rm a,b}$ & \ 8 & 40 & LMC & & \\
&&&&NGC~1795 &  29 & 41 & LMC   & & \\* 

SWB VI\dotfill& 6$\times$10$^9$& 0.004& 0\arcsec --- 60\arcsec & NGC~1751&  
44 & 42&    LMC&  0.12 & \\   
&&&&NGC~2173 & 31 & 42 &  LMC &    0.07    &                  V-VI \\
&&&&NGC~1652$^{\rm a}$ & 15 & 43 & LMC & & \\
&&&&ESO121-SCO3$^{\rm b}$& \ 9 & 44&LMC &            &                  \\
&&&&NGC~2121 & 47 & 44&  LMC &    0.10    &                  VI   \\
&&&&NGC~1718 & 33 & 45 & LMC &            &        \\   
&&&&NGC~1978 & 28 & 45 & LMC &    0.10    &                  VI\\
&&&&NGC~1852 & 33 & 45 &  LMC &            &   \\
&&&&NGC~2155 & 18  & 45 & LMC &    0.10    &                  VI\\
&&&&SL~842$^{\rm b}$  &\ 5 & 45 &LMC   &            &       \\
&&&&L~1$^{\rm a}$ & 26 & 46 & SMC & & \\
&&&&NGC~416  & 25 & 46&  SMC  &            &             VI \\
&&&&NGC~1754 & 25 & 46&   LMC &            &           \\
&&&&NGC~1916$^{\rm a}$ & 16 & 46 & LMC & & \\
&&&&NGC~2005  & 43 & 46&   LMC &            &         \\  
&&&&NGC~2019  & 46 & 46&   LMC &      0.18  &                    VII\\
&&&&SL~506  & \ 8 & 46  &LMC   &            &          \\

SWB VII\dotfill& 1.2$\times$10$^{10}$& 0.0006 & 0\arcsec --- 60\arcsec & L~11 &  \ 6 & 47 &  SMC    &           &\\
&&&&L~68  & 19 & 47 &  SMC    &           &\\
&&&&NGC~121& 28 &  47&  SMC   &           &             VII\\
&&&&NGC~1835 & 48 & 47 & LMC & 0.12& VII\\
&&&&L~8     & 24 &48  &SMC     &           &                  \\
&&&&NGC~361 & 17 & 48&  SMC   &           & \\
&&&&NGC~1786& 35 & 48&   LMC  &     0.12  &            \\
&&&&NGC~2210& 31 & 48&  LMC  &  0.10     &                 VII\\
&&&&L~113   & 15 &49  & SMC    &           &\\
&&&&NGC~339 & 22 & 49&  SMC   &           &            VII\\
&&&&NGC~1898$^{\rm a}$& 20 & 50 & LMC & 0.09 & \\
&&&&H~11 (SL~868)    & 20 & 51  & LMC    &   0.10    &                  VII\\

\tablecomments
{Col.\ (2) and (3). Ages and metallicities of superclusters 
from \citet{cohe82}; for the Pre-SWB supercluster only, we have 
adopted the age of a cluster with $s$ = 7 from \citet{elso85}. Col.\ (5). Number of stars from the 2MASS
PSC included in analysis. 
Col.\ (7). $s$-parameter from \citet{elso85,elso88}. 
Col.\ (9). $E(B - V)$\ \ from \citet{pers83};\ \ 
otherwise,\ \ we\ \ have\ \ assumed\ \ $E(B - V)$ = 0.075,\ \ and $E(B - V)$ = 0.037
for the SMC \citep{schl98}. Col.\ (10). SWB class from \citet{sear80}.\\
\\
\noindent
$^{\rm a}$\ \ Data from All Sky Data Release of 2MASS.\\  
\noindent 
\\
$^{\rm b}$\ \ Clusters added to sample from \citet{elso88}.\\ 
\\
\noindent
$^{\rm c}$\ \ Listed as IC~1665 in both \citet{vand81}, 
   and \citet{elso85}. 
}

\enddata
\label{tabclust}
\end{deluxetable}
\end{landscape}

%
%
%
%
%
%
%
%
%
%

\begin{landscape}
\begin{deluxetable}{lcllccccc}
\tabletypesize{\small}
\tablewidth{8.25in}
\tablecaption{Fornax Cluster Galaxies}
\tablehead{
\colhead{Name} &
\colhead{\barM$^{I-SBF}_{K_s}$} &
\colhead{\barM$^{I-SBF}_{F160W}$} &
\colhead{\barM$_H$} &
\colhead{$J$ } &
\colhead{$H$ } &
\colhead{$K_s$ } &
\colhead{Age (Gyr)} &
\colhead{[Fe/H]} \\[-0.3cm]
}
\startdata

IC~2006&            -5.83 $\pm$ 0.31& -4.82 $\pm$ 0.29 & -5.09 $\pm$ 0.31&\hspace*{0.18cm}9.575$\pm$ 0.019&  \hspace*{0.18cm}8.866$\pm$ 0.018& \hspace*{0.18cm}8.638$\pm$ 0.013& \hspace*{0.18cm}6.0$\pm$2.1 &  \hspace*{0.18cm}0.25$\pm$0.15\\
NGC~1336&           -5.67 $\pm$ 0.31& \hspace*{0.86cm}... &\hspace*{0.86cm}... & 10.948$\pm$ 0.022&10.255$\pm$ 0.020&10.031$\pm$ 0.028& \hspace*{0.18cm}9.0$\pm$3.1 & -0.12$\pm$0.15\\
NGC~1339&           -5.76 $\pm$ 0.36& -5.02 $\pm$ 0.36 & -5.28 $\pm$ 0.37& \hspace*{0.18cm}9.665$\pm$ 0.018& \hspace*{0.18cm}8.998$\pm$ 0.012 & \hspace*{0.18cm}8.771$\pm$ 0.014&\hspace*{0.18cm}8.0$\pm$2.8 & \hspace*{0.18cm}0.15$\pm$0.15\\
NGC~1351&           -5.68 $\pm$ 0.19& -4.77 $\pm$ 0.17 & -5.02 $\pm$ 0.20&   \hspace*{0.18cm}9.610$\pm$ 0.018& \hspace*{0.18cm}8.931$\pm$ 0.013 & \hspace*{0.18cm}8.745$\pm$ 0.015& 10.5$\pm$3.6 &  \hspace*{0.18cm}0.00$\pm$0.15\\
NGC~1373&           -6.67 $\pm$ 0.60& -5.2 \hspace*{0.18cm}$\pm$ 0.5 & -5.5 \hspace*{0.18cm}$\pm$ 0.5& 11.538$\pm$ 0.021& 10.878$\pm$ 0.021& 10.693$\pm$ 0.028&  \hspace*{0.18cm}9.5$\pm$3.3 & \hspace*{0.18cm}0.00$\pm$0.15\\
NGC~1374&           -5.82 $\pm$ 0.14& -4.82 $\pm$ 0.18 & -5.08 $\pm$ 0.21 &  \hspace*{0.18cm}9.134$\pm$ 0.017& \hspace*{0.18cm}8.446$\pm$ 0.012& \hspace*{0.18cm}8.242$\pm$  0.012 & 11.0$\pm$3.8&  \hspace*{0.18cm}0.05$\pm$0.15\\
NGC~1375&           -6.06 $\pm$ 0.29& -5.47 $\pm$ 0.15 & -5.72 $\pm$ 0.18 &  10.635$\pm$ 0.019&10.000$\pm$ 0.017& \hspace*{0.18cm}9.766$\pm$ 0.021 & \hspace*{0.18cm}1.5$\pm$0.5 &  \hspace*{0.18cm}0.30$\pm$0.15\\
NGC~1379&           -5.85 $\pm$ 0.17& -5.11 $\pm$ 0.19 & -5.36 $\pm$ 0.21 & \hspace*{0.18cm}9.263$\pm$ 0.018& \hspace*{0.18cm}8.606$\pm$ 0.012 & \hspace*{0.18cm}8.396$\pm$ 0.013  & \hspace*{0.18cm}8.0$\pm$2.8 &  -0.02$\pm$0.15\\
NGC~1380&           -5.84 $\pm$ 0.18& -4.64 $\pm$ 0.19 & -4.91 $\pm$ 0.22 &   \hspace*{0.18cm}8.042$\pm$ 0.017& \hspace*{0.18cm}7.355$\pm$ 0.011 &\hspace*{0.18cm}7.092$\pm$ 0.010   & \hspace*{0.18cm}6.3$\pm$2.2 & \hspace*{0.18cm}0.28$\pm$0.15\\
NGC~1404&           -5.72 $\pm$ 0.20& -4.76 $\pm$ 0.21 & -5.03 $\pm$ 0.23 &  \hspace*{0.18cm}7.838$\pm$ 0.017& \hspace*{0.18cm}7.149$\pm$ 0.011 &\hspace*{0.18cm}6.902$\pm$ 0.010   & \hspace*{0.18cm}5.0$\pm$1.7 &  \hspace*{0.18cm}0.30$\pm$0.15\\
NGC~1427&           -6.40 $\pm$ 0.25& -5.28 $\pm$ 0.25 & -5.53 $\pm$ 0.27 &  \hspace*{0.18cm}9.089$\pm$ 0.019& \hspace*{0.18cm}8.399$\pm$ 0.018 &\hspace*{0.18cm}8.223$\pm$ 0.012   & \hspace*{0.18cm}7.0$\pm$2.4 &  \hspace*{0.18cm}0.15$\pm$0.15\\

\tablecomments{
Col.\hspace*{0.18cm}(2). Absolute $K_s$ fluctuation magnitude from
\citet{liu02}. Col.\hspace*{0.18cm}(3). Absolute 
$F160W$ fluctuation magnitude from \citet{jens03}. 
Col.\hspace*{0.18cm}(3). Absolute $H$ fluctuation magnitude calculated 
from $F160W$ values via the transformations in \citet{step00} (see text).
Col.\hspace*{0.18cm}(5), (6), and (7). $J$, $H$, and $K_s$ magnitudes 
from 2MASS Extended Source Catalog (XSC). Col.\hspace*{0.18cm}(8) and (9). Ages and
abundances from \citet{kunt98}; errors of $\pm$0.15 dex 
in ages and abundances
are averages taken from Figure 12 in \citet{kunt00}. 
}


\enddata
\label{tabfornax}
\end{deluxetable}
\end{landscape}

\begin{landscape}
\begin{deluxetable}{lcllccccccc}
\tabletypesize{\scriptsize}
\tablewidth{8.4in}
\tablecaption{Results I: Supercluster integrated magnitudes and colors}
\tablehead{
\colhead{Supercluster} &
\colhead{$N_{\rm cl}$} &
\colhead{Age (yr)} &
\colhead{Z} &
\colhead{Mass ($10^6$ M$_\odot$)} &
\colhead{$M_J$} &
\colhead{$M_H$} &
\colhead{$M_{K_s}$} &
\colhead{($J - H$)} &
\colhead{($J - K_s$)} &
\colhead{($H - K_s$)} \\[-0.3cm] 
}

\startdata

Pre-SWB \dotfill & 28 &2.4$\pm1.7\times$10$^6$     & 0.010\hspace*{0.15cm}$\pm$0.005 &\hspace*{-0.13cm}1.1\hspace*{0.15cm}$\pm$0.4 &  -12.79$\pm$0.09& -13.23$\pm$0.10& -13.58$\pm$0.09&   0.47$\pm$0.04&   0.81$\pm$0.04&   0.34$\pm$0.03 \\

SWB I \dotfill   & 29 &1.0$\pm0.7\times$10$^7$       & 0.010\hspace*{0.15cm}$\pm$0.005 &0.53$\pm$0.06&    -13.59$\pm$0.07& -14.22$\pm$0.07& -14.45$\pm$0.07&   0.63$\pm$0.03&   0.86$\pm$0.03&   0.23$\pm$0.03 \\

SWB II \dotfill  &35 & 5\hspace*{0.24cm}$\pm3\hspace*{0.21cm}\times$10$^7$       & 0.010\hspace*{0.15cm}$\pm$0.005 &0.96$\pm$0.09            &    -13.12$\pm$0.07& -13.69$\pm$0.07& -13.85$\pm$0.08&   0.57$\pm$0.03&   0.72$\pm$0.04&   0.15$\pm$0.03 \\

SWB III \dotfill & 32 &5\hspace*{0.24cm}$\pm3\hspace*{0.21cm}\times$10$^8$       & 0.010\hspace*{0.15cm}$\pm$0.005 &\hspace*{-0.13cm}2.0\hspace*{0.15cm}$\pm$0.1                    &  -12.67$\pm$0.06& -13.18$\pm$0.08& -13.37$\pm$0.08&   0.51$\pm$0.04&   0.70$\pm$0.04&   0.18$\pm$0.03 \\

SWB IV \dotfill &18 &1.0$\pm0.7\times$10$^9$       & 0.010\hspace*{0.15cm}$\pm$0.005 &\hspace*{-0.13cm}2.0\hspace*{0.15cm}$\pm$0.2                                              &   -11.90$\pm$0.07& -12.53$\pm$0.09& -12.73$\pm$0.09&   0.64$\pm$0.04&   0.83$\pm$0.05&   0.19$\pm$0.04 \\

SWB V \dotfill   & 20 &3\hspace*{0.24cm}$\pm2\hspace*{0.21cm}\times$10$^9$       & 0.008\hspace*{0.15cm}$\pm$0.004 &\hspace*{-0.13cm}4.2\hspace*{0.15cm}$\pm$0.2                    &    -12.41$\pm$0.05& -13.07$\pm$0.05& -13.33$\pm$0.06&   0.66$\pm$0.03&   0.92$\pm$0.04&   0.26$\pm$0.03 \\

SWB VI \dotfill &17 &6\hspace*{0.24cm}$\pm4\hspace*{0.21cm}\times$10$^9$       & 0.004\hspace*{0.15cm}$\pm$0.002 &\hspace*{-0.13cm}5.8\hspace*{0.15cm}$\pm$0.1                   &  -12.37$\pm$0.05& -12.97$\pm$0.05& -13.14$\pm$0.06&   0.61$\pm$0.03&   0.78$\pm$0.04&   0.17$\pm$0.03 \\

SWB VII \dotfill &12 & 1.2$\pm0.8\times$10$^{10}$ & 0.0006$\pm$0.0003 & \hspace*{-0.13cm}7.6\hspace*{0.15cm}$\pm$0.1                                                              &   -12.14$\pm$0.05& -12.66$\pm$0.05& -12.74$\pm$0.05&   0.52$\pm$0.03&   0.60$\pm$0.03&   0.08$\pm$0.03 \\

\tablecomments
{Col.\ (2). Number of clusters in each supercluster. Col.\ (3) 
and (4). Ages and metallicities of superclusters, and their errors,
from \citet{cohe82}. Col.\ (5) Masses from  
theoretical near-IR mass--to--light
ratios; errors are equal to the dispersion
of the results at $J$, $H$, and $K_s$.
}

\enddata 
\label{tabresI}
\end{deluxetable}
\end{landscape}

\begin{landscape}
\begin{deluxetable}{lcllccccccc}
\tabletypesize{\scriptsize}
\tablewidth{8.2in}
\tablecaption{Results II: Supercluster fluctuation magnitudes and colors}
\tablehead{
\colhead{Supercluster} &
\colhead{$N_{\rm cl}$} &
\colhead{Age (yr)} &
\colhead{Z} &
\colhead{Mass ($10^6$ M$_\odot$)} &
\colhead{\barM$_J$} &
\colhead{\barM$_H$} &
\colhead{\barM$_{K_s}$} &
\colhead{(\barJ\ - \barH)} &
\colhead{(\barJ\ - \barKs)} &
\colhead{(\barH\ - \barKs)}\\[-0.3cm]
}

\startdata

Pre-SWB \dotfill & 28 &2.4$\pm1.7\times$10$^6$     & 0.010\hspace*{0.15cm}$\pm$0.005 &\hspace*{-0.13cm}1.1\hspace*{0.15cm}$\pm$0.4 &  -6.49$\pm$0.42&  -7.63$\pm$0.43&  -7.70$\pm$0.40&   1.14$\pm$0.07&   1.22$\pm$0.09&   0.07$\pm$0.05 \\

SWB I \dotfill   &29&1.0$\pm0.7\times$10$^7$       & 0.010\hspace*{0.15cm}$\pm$0.005 &0.53$\pm$0.06&   -7.54$\pm$0.14&  -8.49$\pm$0.12&  -8.67$\pm$0.12&   0.95$\pm$0.06&   1.13$\pm$0.06&   0.18$\pm$0.01 \\

SWB II \dotfill  &35&5\hspace*{0.24cm}$\pm3\hspace*{0.21cm}\times$10$^7$       & 0.010\hspace*{0.15cm}$\pm$0.005 &0.96$\pm$0.09                 &  -6.74$\pm$0.40&  -7.57$\pm$0.32&  -7.88$\pm$0.28&   0.84$\pm$0.10&   1.14$\pm$0.14&   0.30$\pm$0.05 \\

SWB III \dotfill &32&5\hspace*{0.24cm}$\pm3\hspace*{0.21cm}\times$10$^8$       & 0.010\hspace*{0.15cm}$\pm$0.005 &\hspace*{-0.13cm}2.0\hspace*{0.15cm}$\pm$0.1                 &  -6.05$\pm$0.23 &  -7.10$\pm$0.23&-7.45$\pm$0.24  &  1.05$\pm$0.04&   1.39$\pm$0.05& 0.35$\pm$0.02 \\

SWB IV \dotfill  &18&1.0$\pm0.7\times$10$^9$       & 0.010\hspace*{0.15cm}$\pm$0.005 &\hspace*{-0.13cm}2.0\hspace*{0.15cm}$\pm$0.2                                           &  -5.67$\pm$0.22&  -6.85$\pm$0.21&  -7.51$\pm$0.18&   1.18$\pm$0.05&   1.84$\pm$0.12&   0.66$\pm$0.09  \\

SWB V \dotfill   &20&3\hspace*{0.24cm}$\pm2\hspace*{0.21cm}\times$10$^9$       & 0.008\hspace*{0.15cm}$\pm$0.004 &\hspace*{-0.13cm}4.2\hspace*{0.15cm}$\pm$0.2                 &   -4.60$\pm$0.14&  -5.94$\pm$0.16&  -6.69$\pm$0.20&   1.34$\pm$0.05&   2.09$\pm$0.10&   0.76$\pm$0.06  \\

SWB VI \dotfill  &17&6\hspace*{0.24cm}$\pm4\hspace*{0.21cm}\times$10$^9$       & 0.004\hspace*{0.15cm}$\pm$0.002 & \hspace*{-0.13cm}5.8\hspace*{0.15cm}$\pm$0.1                 &  -4.23$\pm$0.17 & -5.49$\pm$0.19 &  -6.21$\pm$0.24 & 1.26$\pm$0.07&   1.98$\pm$0.16 &  0.72$\pm$0.10\\

SWB VII \dotfill &12&1.2$\pm0.8\times$10$^{10}$ & 0.0006$\pm$0.0003 & \hspace*{-0.13cm}7.6\hspace*{0.15cm}$\pm$0.1                                                           &  -3.14$\pm$0.19&  -4.29$\pm$0.27&  -4.92$\pm$0.38&   1.15$\pm$0.10&   1.78$\pm$0.22&   0.63$\pm$0.14  \\

\enddata
\label{tabresII}
\end{deluxetable}
\end{landscape}

\begin{deluxetable}{lclclcl}
\tablewidth{5.1in}
\tablecaption{ Fractional contribution of bright stars to integrated light }
\tablehead{
&\multicolumn{2}{c}{$J$}& \multicolumn{2}{c}{$H$} & \multicolumn{2}{c}{$K_s$}\\
\colhead{Supercluster} &
\colhead{data}&
\colhead{model}&
\colhead{data}&
\colhead{model}&
\colhead{data}&
\colhead{model}\\[-0.3cm]
}

\startdata
Pre-SWB \dotfill  &47$\pm$6\%&39$\pm$5\%&67$\pm$9\%&39$\pm$5\%&57$\pm$8\%&38$\pm$6\%\\

SWB I \dotfill    &68$\pm$6\%&85$\pm$7\%&81$\pm$7\%&91$\pm$7\%&78$\pm$6\%&92$\pm$8\%\\

SWB II \dotfill   &61$\pm$4\%&80$\pm$5\%&71$\pm$4\%&87$\pm$5\%&77$\pm$4\%&88$\pm$5\%\\

SWB III \dotfill  &63$\pm$5\%&23$\pm$2\%&75$\pm$6\%&36$\pm$3\%&76$\pm$6\%&43$\pm$3\%\\

SWB IV \dotfill   &55$\pm$6\%&23$\pm$2\%&64$\pm$7\%&36$\pm$4\%&73$\pm$8\%&43$\pm$5\%\\

SWB V \dotfill    &36$\pm$3\%&29$\pm$2\%&45$\pm$4\%&39$\pm$3\%&49$\pm$5\%&44$\pm$4\%\\

SWB VI \dotfill   &33$\pm$3\%&27$\pm$2\%&40$\pm$3\%&35$\pm$3\%&45$\pm$4\%&39$\pm$4\%\\

SWB VII \dotfill  &20$\pm$2\%&20$\pm$2\%&25$\pm$2\%&25$\pm$2\%&27$\pm$3\%&27$\pm$3\%\\

\enddata 
\label{tabfrac}
\end{deluxetable}


\begin{deluxetable}{cccccc}
\tablewidth{3.7in}
\tablecaption{ Fluctuation color vs log (age/yr)  }
\tablehead{
\colhead{$\overline{Color}$} &
\colhead{$a$} &
\colhead{$b$} &
\colhead{$N$} &
\colhead{rms} &
\colhead{$\widetilde{\chi}^2$} 
\\[-0.3cm]    
}

\startdata

($\overline{J} - \overline{H}$)&0.29$\pm$0.18&0.10$\pm$0.02&8&0.37&2.9\\
($\overline{H} - \overline{K_s}$)&-1.29$\pm$0.22&0.21$\pm$0.03&8&0.26&1.2\\
($\overline{H} - \overline{K_s}$)&-1.25$\pm$0.16&0.20$\pm$0.02&18&0.76&1.8\\
($\overline{J} - \overline{K_s}$)&-0.97$\pm$0.36&0.30$\pm$0.04&8&0.56&2.1\\

\tablecomments{Fits of fluctuation color vs age of the form 
$\overline{\rm Color} = a + b\ [\ {\rm log(age/yr)}\ ]$; the number 
of objects used
for the fit is tabulated as $N$. The resulting rms of the points 
(in magnitudes) after the fit and the
reduced chi-square ($\widetilde{\chi}^2$) are also listed. }
\enddata
\label{tabflcol}
\end{deluxetable}


\clearpage

\vspace*{-1cm}
\hspace*{0cm}\epsfig{file=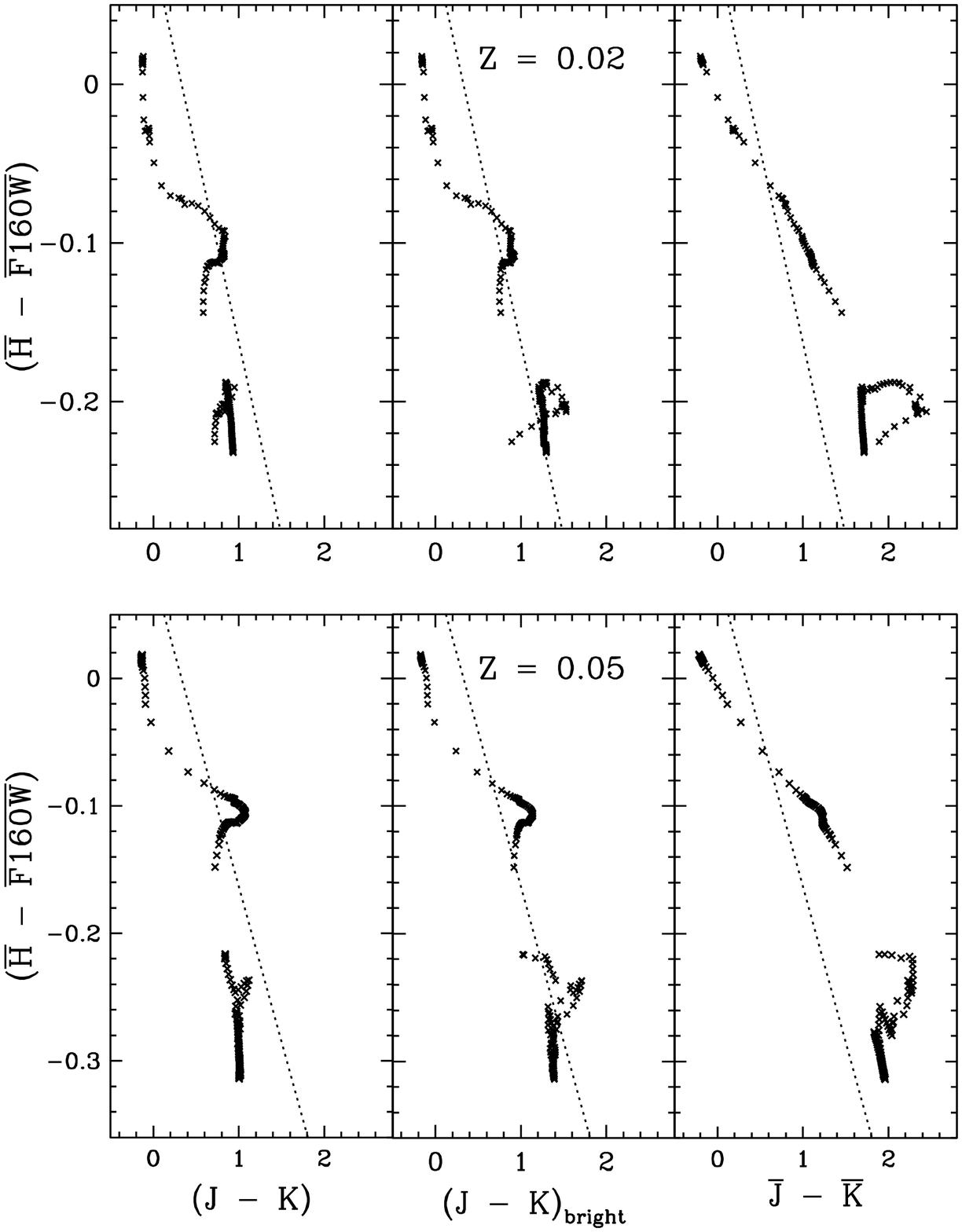,height=1.0\textheight}
\clearpage


\clearpage

\hspace*{-3cm}\epsfig{file=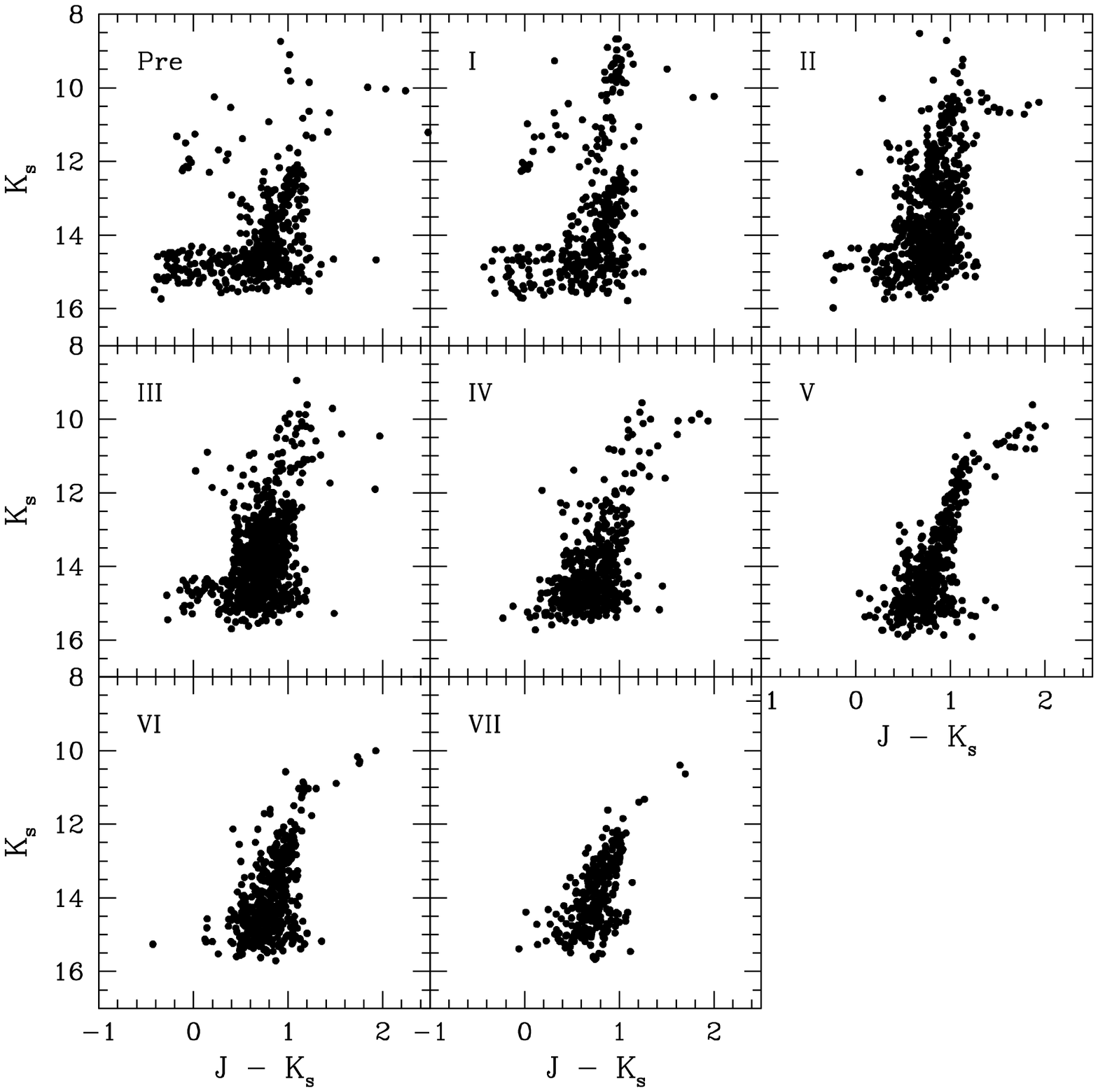, height=0.9\textheight}

\newpage

\vspace*{-2cm}
\hspace*{-2cm}\psfig{file=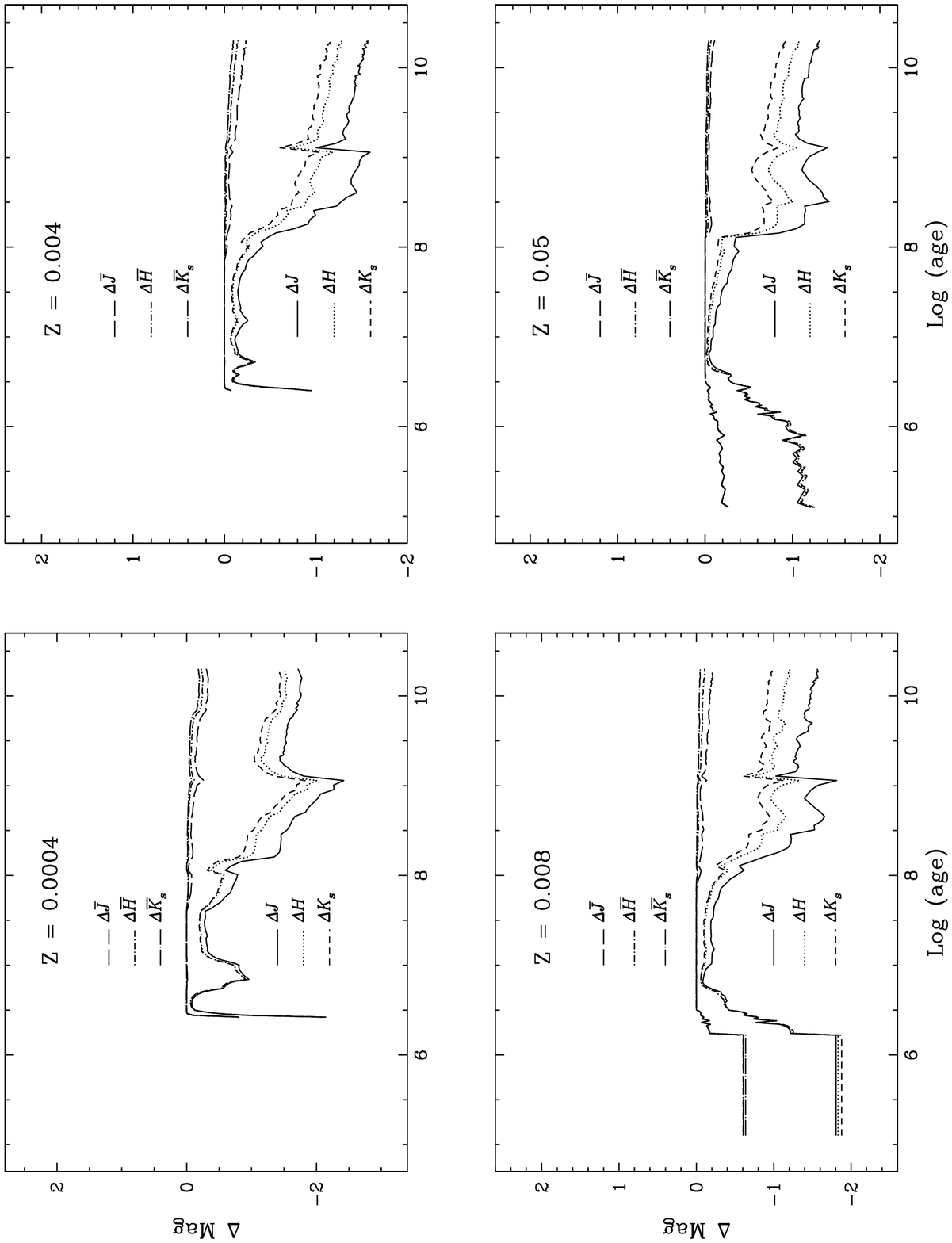,angle=0,width=0.85\textheight}

\newpage

\vspace*{-5cm}
\hspace*{-4.5cm}\epsfig{file=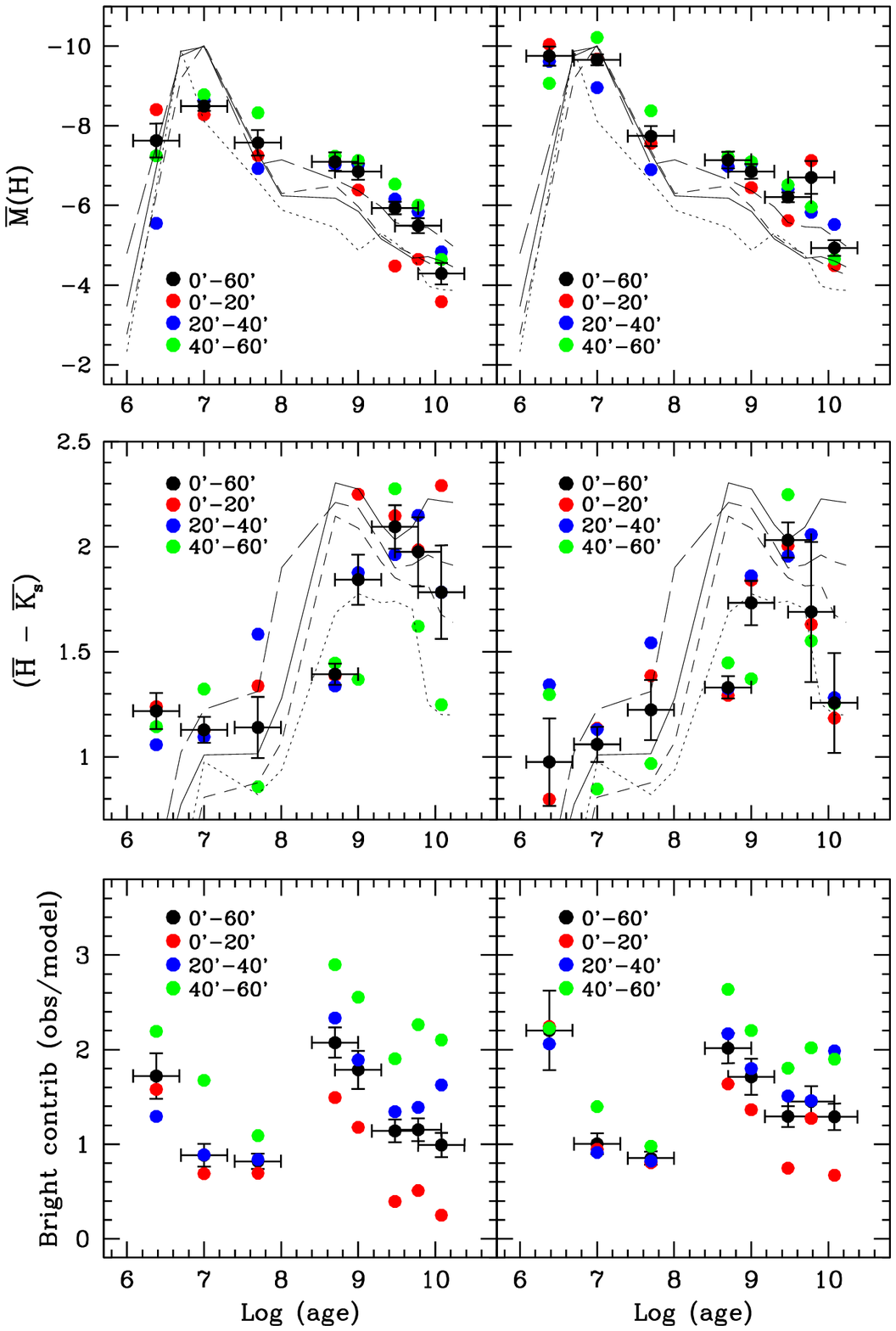,height=1.4\textheight}
\clearpage

\vspace*{-1cm}
\hspace*{-0.5cm}\psfig{file=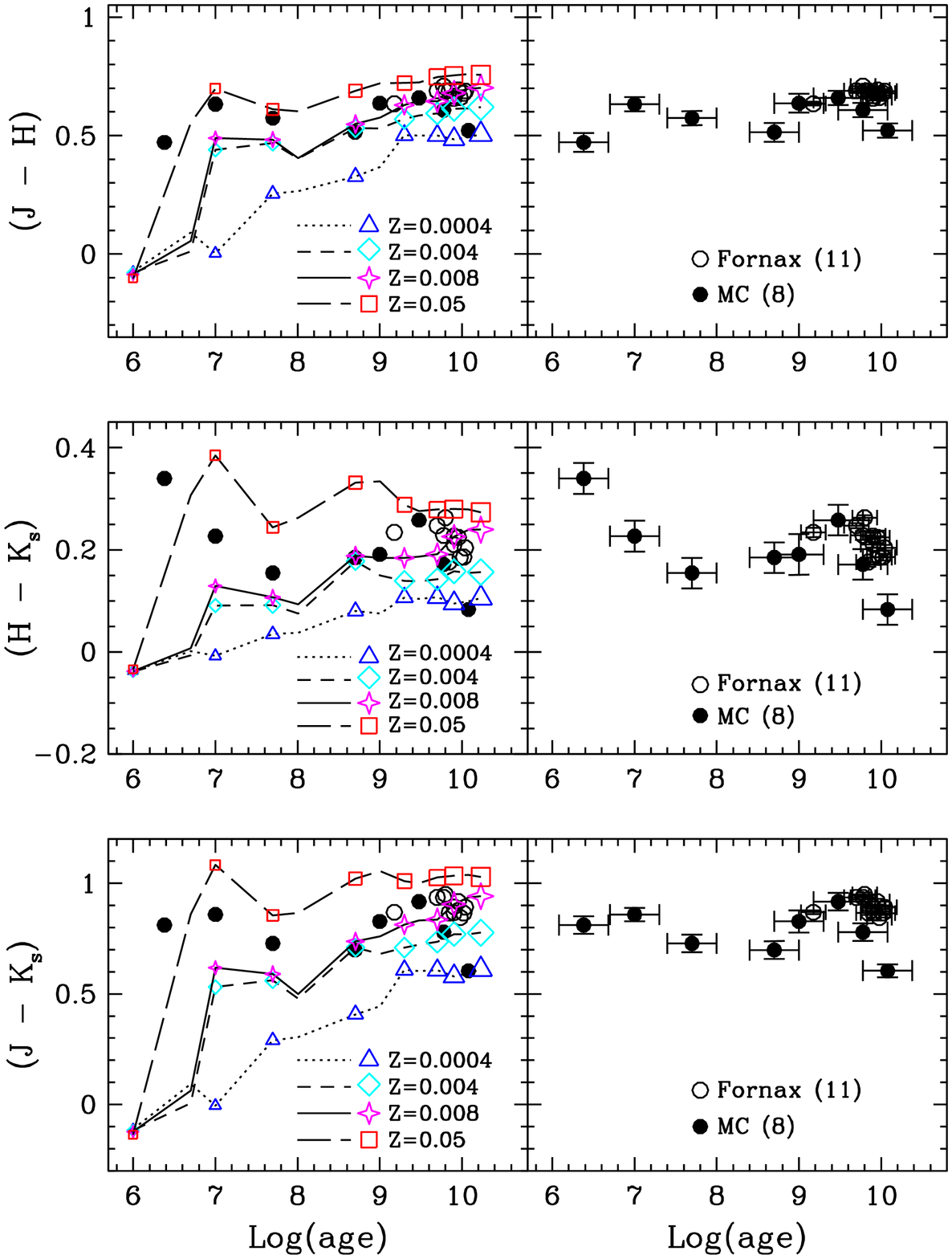,angle=0,width=0.75\textheight}

\clearpage

\vspace*{-5cm}
\hspace*{-4.5cm}\epsfig{file=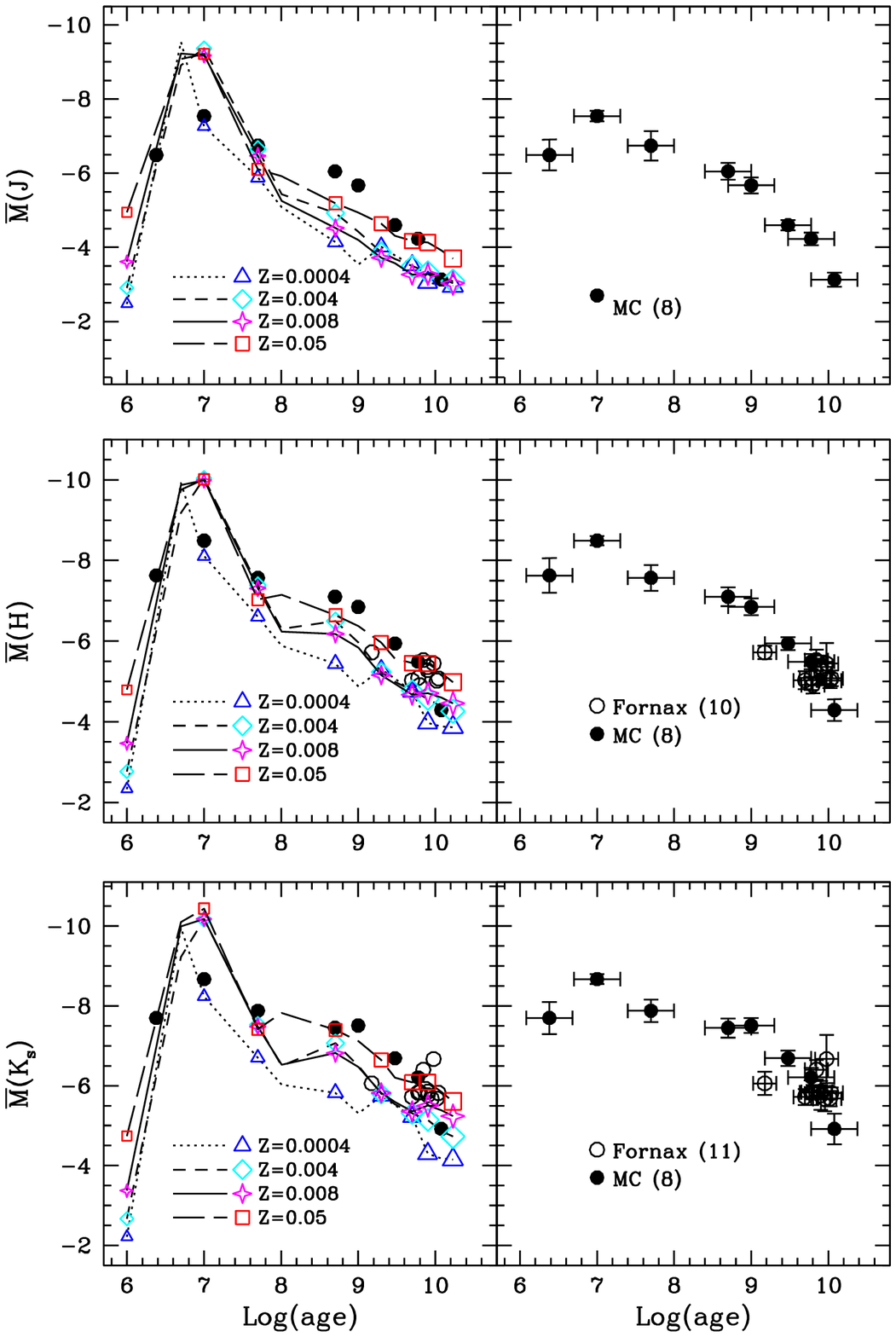,height=1.2\textheight}
\clearpage

\vspace*{0cm}
\hspace*{-0.5cm}\epsfig{file=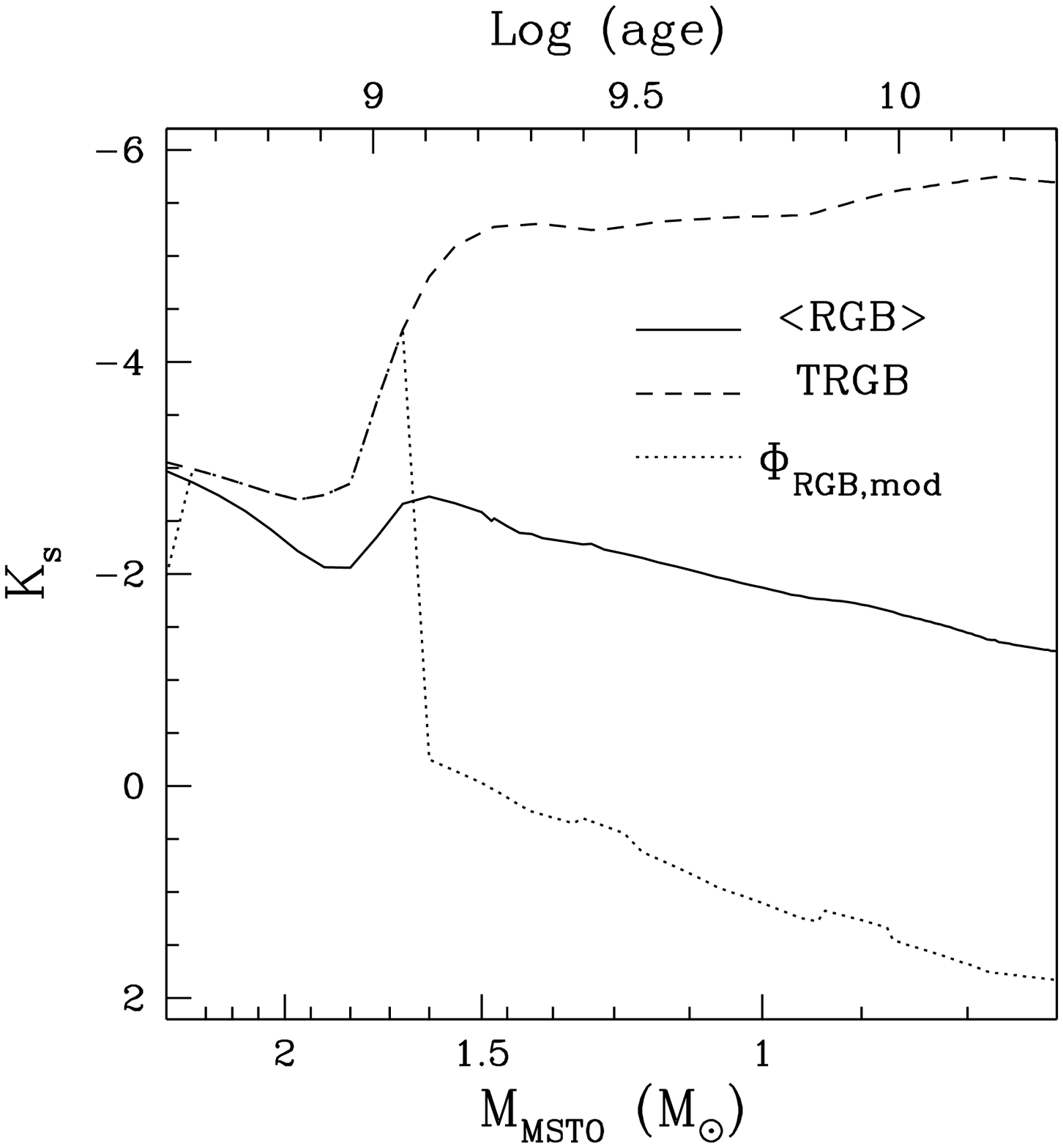,height=0.80\textheight}

\newpage

\vspace*{2.5cm}
\hspace*{-2cm}\epsfig{file=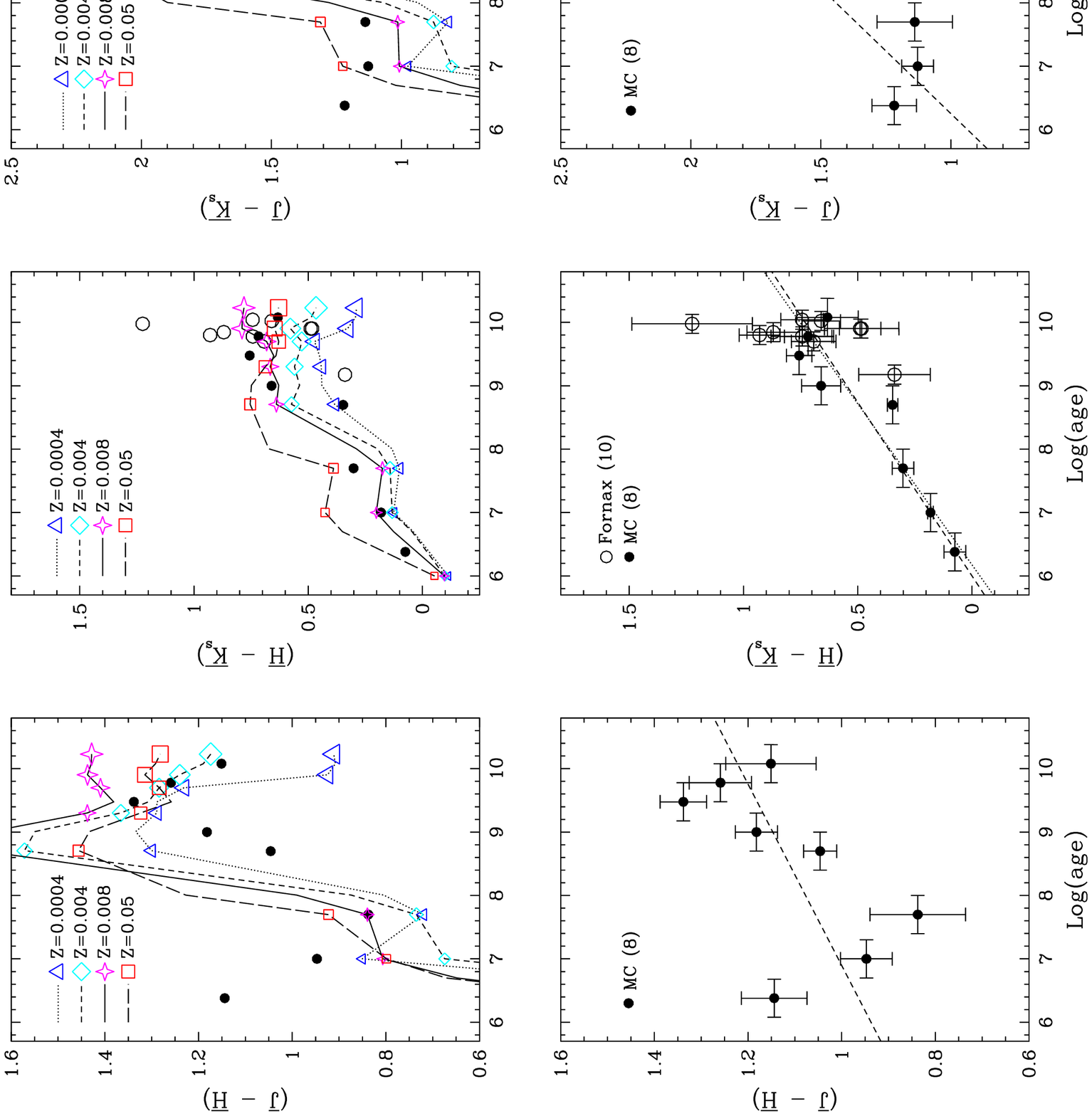,angle=90,height=1.1\textwidth} 
\clearpage

\vspace*{0cm}
\hspace*{-0.5cm}\epsfig{file=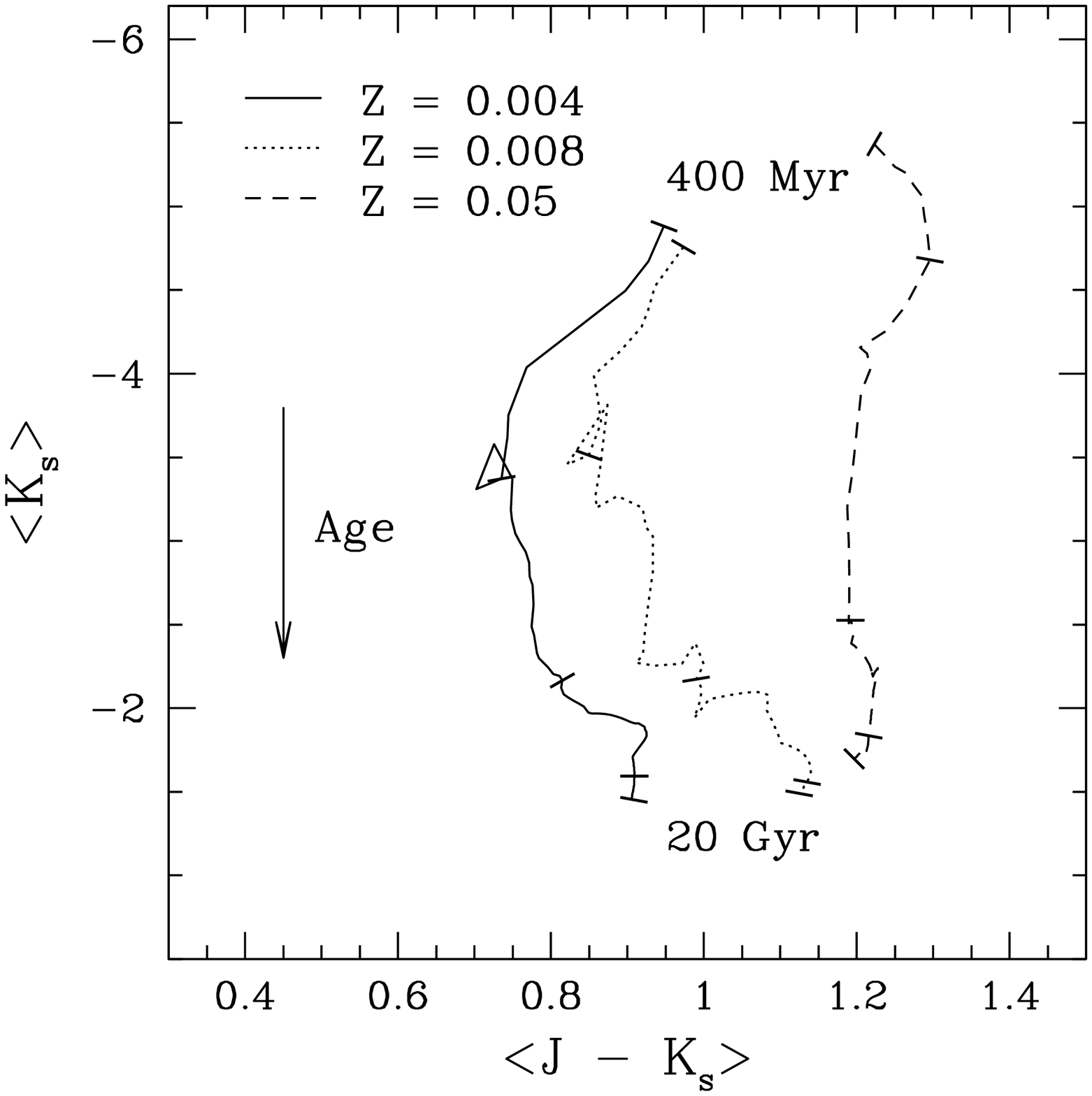,height=0.80\textheight}

\clearpage

\vspace*{-5cm}
\hspace*{-4.5cm}\epsfig{file=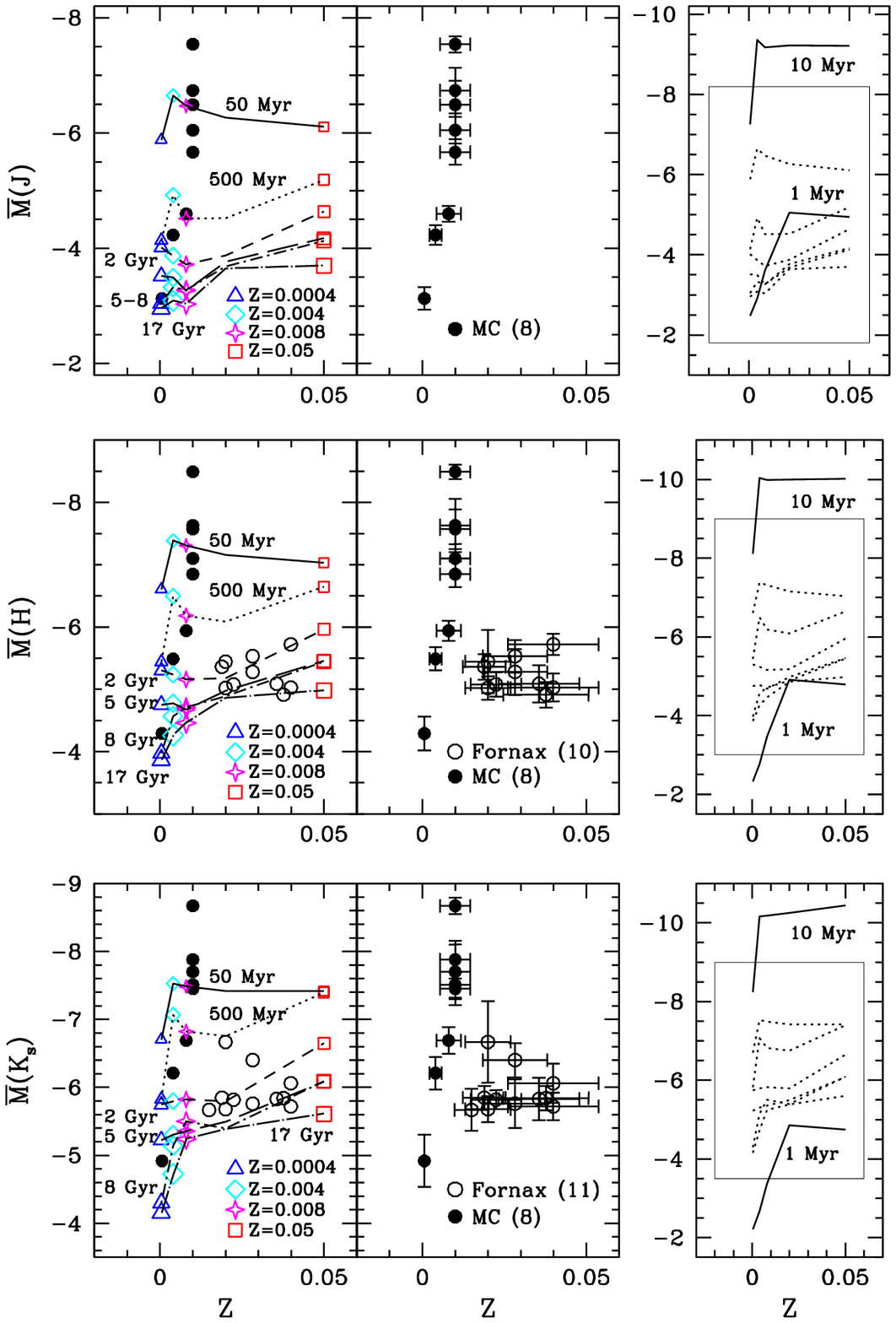,height=1.2\textheight}
\clearpage

\vspace*{0cm}
\hspace*{-1.5cm}\epsfig{file=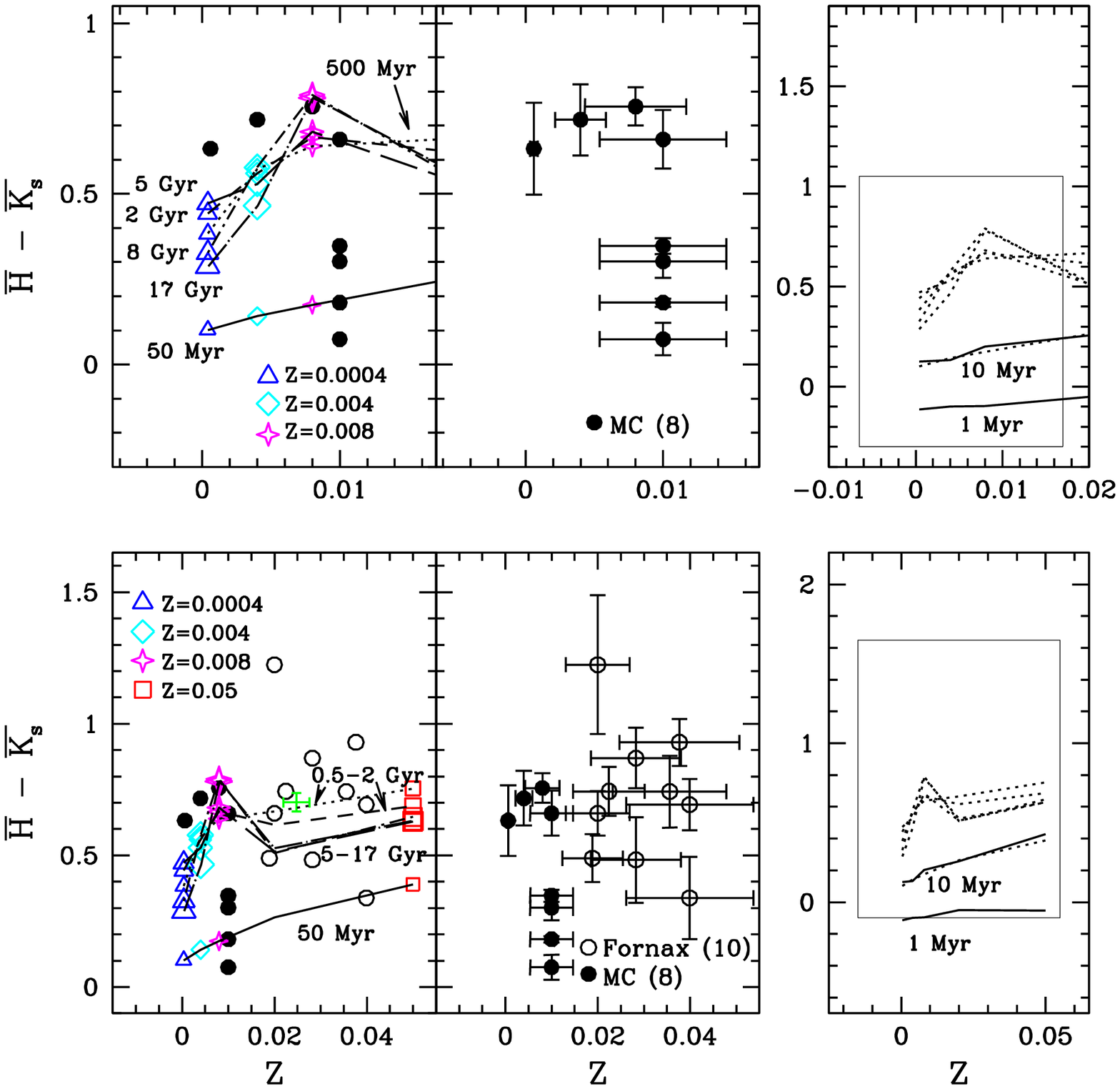,height=0.9\textheight}


\begin{thebibliography}{}


\bibitem[Ajhar \& Tonry(1994)]{ajha94} Ajhar, E.~A.~\& Tonry, 
J.~L.\ 1994, \apj, 429, 557 

\bibitem[Alongi et al.(1993)]{alon93} Alongi, M., Bertelli, 
G., Bressan, A., Chiosi, C., Fagotto, F., Greggio, L., \& Nasi, E.\ 1993, 
\aaps, 97, 851 



\bibitem[Blakeslee, Vazdekis, \& Ajhar(2001)]{blak01} 
Blakeslee, J.~P., Vazdekis, A., \& Ajhar, E.~A.\ 2001, \mnras, 320, 193

\bibitem[Bressan et al.(1993)]{bres93} 
Bressan, A., Fagotto, F., Bertelli, G., \& Chiosi, C.\ 1993, \aaps, 100, 
647 

\bibitem[Bruzual (2002)]{bruz02} Bruzual A., G. 2002, in 
Extragalactic Star Clusters, IAU Symposium Ser., Vol.\ 207, eds.\
D. Geisler, E.~K. Grebel, \& D. Minniti (Provo:ASP), 616  



\bibitem[Bruzual \& Charlot(2003)]{bruz03} Bruzual, A. G., 
    \& Charlot, S. 2003, \mnras, 344, 1000  

\bibitem[Buzzoni(1989)]{buzz89} Buzzoni, A.\ 1989, \apjs, 71, 817 

\bibitem[Buzzoni(1993)]{buzz93} 
\samename 1993, \aap, 275, 433 


\bibitem[Cantiello et al.(2003)]
{cant03} Cantiello, M., Raimondo, G., 
Brocato, E., \& Capaccioli, M.\ 2003, \aj, 125, 2783 

\bibitem[Cervi{\~ n}o et al.\ (2002)Cervi{\~ n}o, Valls--Gabaud, Luridiana, \& 
Mas--Hesse]{cerv02} Cervi{\~ n}o, M., Valls--Gabaud, D., 
Luridiana, V., \& Mas--Hesse, J.~M.\ 2002, \aap, 381, 51 

\bibitem[Cervi{\~ n}o \& Valls--Gabaud(2003)]{cerv03} Cervi{\~ 
n}o, M.~\& Valls--Gabaud, D.\ 2003, \mnras, 338, 481

\bibitem[Chabrier(2003)]{chab03} Chabrier, G.\ 2003, \pasp, 
115, 763 


\bibitem[Charlot \& Fall(2000)]{char00} Charlot, S.~\& 
        Fall, S.~M.\ 2000, \apj, 539, 718 

\bibitem[Charlot, Worthey, \& Bressan(1996)]{char96} Charlot, 
S., Worthey, G., \& Bressan, A.\ 1996, \apj, 457, 625

\bibitem[Cohen(1982)]{cohe82} Cohen, J.~G.\ 1982, \apj, 258, 
143

\bibitem[Cutri et al.(2003)]{cutr03} Cutri, R.~M.\ 2003, 
Explanatory Supplement to the 2MASS All Sky Data
Release, http://www.ipac.caltech.edu/2mass/releases/allsky/doc/explsup.html


\bibitem[Elson \& Fall(1985)]{elso85} Elson, R.~A.~W.~\& Fall, 
S.~M.\ 1985, \apj, 299, 211

\bibitem[Elson \& Fall(1988)]{elso88}  
\samename 1988, \aj, 96, 1383

\bibitem[Faber \& Jackson(1976)]{fabe76} Faber, S.~M.~\&
Jackson, R.~E.\ 1976, \apj, 204, 668

\bibitem[Fagotto et al.(1994a)]{fa94a} 
Fagotto, F., Bressan, A., Bertelli, G., \& Chiosi, C.\ 1994, \aaps, 104, 
365 

\bibitem[Fagotto et al.(1994b)]{fa94b} 
\samename 1994, \aaps, 105, 29 

\bibitem[Ferrarese et al.(2000)]{ferr00} Ferrarese, L.~et al.\ 
2000, \apjs, 128, 431 

\bibitem[Ferraro et al.(1995)]{ferr95} Ferraro, F.~R., Fusi 
Pecci, F., Testa, V., Greggio, L., Corsi, C.~E., Buonanno, R., Terndrup, 
D.~M., \& Zinnecker, H.\ 1995, \mnras, 272, 391

\bibitem[Ferraro et al.(2000)]
{ferra00} Ferraro, F.~R., Montegriffo, P., Origlia, 
L., \& Fusi Pecci, F.\ 2000, \aj, 119, 1282 

\bibitem[Frogel, Persson, \& Cohen(1980)]{frog80} Frogel, 
J.~A., Persson, S.~E., \& Cohen, J.~G.\ 1980, \apj, 239, 495 


\bibitem[Girardi et al.(1996)]{gira96} Girardi, L., Bressan, 
A., Chiosi, C., Bertelli, G., \& Nasi, E.\ 1996, \aaps, 117, 113 


\bibitem[Graham et al.(1998)]{grah98} Graham, A.~W., Colless, 
M.~M., Busarello, G., Zaggia, S., \& Longo, G.\ 1998, \aaps, 133, 325 

%

\bibitem[Harris(1996)]{harr96} Harris, W.~E.\ 1996, \aj, 112, 
1487 

\bibitem[Iben(1967)]{iben67} Iben, I.~J.\ 1967, \araa, 5, 571 


\bibitem[Jensen et al.(2003)]{jens03} Jensen, J.~B., Tonry, 
J.~L., Barris, B.~J., Thompson, R.~I., Liu, M.~C., Rieke, M.~J., Ajhar, 
E.~A., \& Blakeslee, J.~P.\ 2003, \apj, 583, 712


\bibitem[Kuchinski \& Frogel(1995)]{kuch95} Kuchinski, 
L.~E.~\& Frogel, J.~A.\ 1995, \aj, 110, 2844 

\bibitem[Kuntschner(1998)]{kunt98} Kuntschner, H.\ 1998, Ph.D. thesis, 
University of Durham, UK. 

\bibitem[Kuntschner(2000)]{kunt00} \samename 2000, 
\mnras, 315, 184



\bibitem[Lejeune, Cuisinier, \& Buser(1997)]{leje97} 
 Lejeune, T., Cuisinier, F., \& Buser, R.\ 1997, \aaps, 125, 229 

\bibitem[Lejeune, Cuisinier, \& Buser(1998)]{leje98}
\samename 1998, \aaps, 130, 65 

\bibitem[Liu, Charlot, \& Graham(2000)]{liu00} Liu, M.~C., 
Charlot, S., \& Graham, J.~R.\ 2000, \apj, 543, 644 

\bibitem[Liu \& Graham(2001)]{liu01} Liu, M.~C.~\& Graham, 
J.~R.\ 2001, \apjl, 557, L31 

\bibitem[Liu, Graham, \& Charlot(2002)]{liu02} Liu, M.~C., 
Graham, J.~R., \& Charlot, S.\ 2002, \apj, 564, 216 

\bibitem[Mould et al.(2000)]{moul00} Mould, J.~R.~et al.\ 
2000, \apj, 529, 786 

\bibitem[Ochsenbein, Bauer, \& Marcout(2000)]{ochs00} 
Ochsenbein, F., Bauer, P., \& Marcout, J.\ 2000, \aaps, 143, 23 




\bibitem[Persson et al.(1983)]{pers83} Persson, S.~E., 
Aaronson, M., Cohen, J.~G., Frogel, J.~A., \& Matthews, K.\ 1983, \apj, 
266, 105

\bibitem[Prugniel \& Simien(1996)]{prug96} Prugniel, P.~\& 
Simien, F.\ 1996, \aap, 309, 749

\bibitem[Raimondo et al.(2003)]{raim03} Raimondo, G., Brocato, E., 
Cantiello, M. \& Capaccioli, M. 2003, ``A New Theoretical 
Approach to Evaluate Surface Brightness Fluctuations
in Stellar Systems," poster paper in Stellar Populations 2003,
conference, Garching, Germany, October 6--10, 2003 

\bibitem[Renzini(2003)]{renz03} Renzini, A. 2003, personal 
communication

%
\bibitem[Rood \& Crocker (1997)]{rood97} Rood, R.~T., \&
Crocker, D.~A., ``Some Random Unpublished Results of Possible
Interest," http://www.astro.virgina.edu/$\sim$rtr/papers/ 

\bibitem[Santos \& Frogel(1997)]{sant97} Santos, J.~F.~C.~\& 
Frogel, J.~A.\ 1997, \apj, 479, 764 

\bibitem[Schlegel et al.\ (1998)Schlegel, Finkbeiner, \& Davis]{schl98} 
       Schlegel, D.~J., Finkbeiner, D.~P., \& Davis, M.\ 1998, \apj, 500, 525

\bibitem[Searle, Wilkinson, \& Bagnuolo(1980)]{sear80} Searle, 
L., Wilkinson, A., \& Bagnuolo, W.~G.\ 1980, \apj, 239, 803 

\bibitem[Skrutskie et al.(1997)]{skru97} Skrutskie, M.~F.~et 
al.\ 1997, ASSL Vol.~210: The Impact of Large Scale Near--IR Sky Surveys, 25 

\bibitem[Stephens et al.(2000)]{step00} Stephens, A.~W., 
Frogel, J.~A., Ortolani, S., Davies, R., Jablonka, P., Renzini, A., \& 
Rich, R.~M.\ 2000, \aj, 119, 419 

\bibitem[Stetson(1987)]{stet87} Stetson, P.~B.\ 1987, \pasp, 
99, 191


\bibitem[Sweigart \& Gross(1978)]{swei78} Sweigart, A.~V.~\& 
Gross, P.~G.\ 1978, \apjs, 36, 405 

\bibitem[Tonry et al.(1997)]{tonr97} 
Tonry, J.~L., Blakeslee, J.~P., Ajhar, E.~A., \& Dressler, A.\ 1997, \apj, 
475, 399 

\bibitem[Tonry \& Schneider(1988)]{tonr88} Tonry, J.~\& Schneider, 
         D.~P.\ 1988, \aj, 96, 807

\bibitem[van den Bergh(1981)]{vand81} van den Bergh, S.\ 1981, 
\aaps, 46, 79

\bibitem[VandenBerg \& Bell(1985)]{vand85} VandenBerg, 
D.~A.~\& Bell, R.~A.\ 1985, \apjs, 58, 561 

\bibitem[Welch(1991)]{welc91} Welch, D.~L.\ 1991, \aj, 101, 
538 

\bibitem[Westera(2001)]{west01} Westera, P.\ 2001, Ph.D.
Thesis, University of Basel

\bibitem[Westera et al.(2002)]{west02} Westera, P., Lejeune, 
T., Buser, R., Cuisinier, F., \& Bruzual, G.\ 2002, \aap, 381, 524 

\bibitem[Worthey(1993a)]{wort93a} Worthey, G.\ 1993a, \apj, 409, 
530

\bibitem[Worthey(1993b)]{wort93b} \samename 1993b, \apj, 418, 
947

\end{thebibliography}
\end{document}